\documentclass[preprint,pre,aps,amsfonts,amsmath,showpacs,superscriptaddress,nofootinbib]{revtex4}
\usepackage{amsmath,amssymb,amscd}
\usepackage{graphicx}
\usepackage{bm}
\begin{document}
\title{On the Gap and Time Interval between the First Two Maxima of Long Random Walks}
\author{Satya N. Majumdar}
\email{majumdar@lptms.u-psud.fr}
\affiliation{Laboratoire de Physique Th\'eorique et Mod\`eles Statistiques, Universit\'e Paris-Sud, B\^at 100, 91405 Orsay Cedex, France.}
\author{Philippe Mounaix}
\email{philippe.mounaix@cpht.polytechnique.fr}
\affiliation{Centre de Physique Th\'eorique, UMR 7644 du CNRS, Ecole
Polytechnique, 91128 Palaiseau Cedex, France.}
\author{Gr\'egory Schehr}
\email{gregory.schehr@lptms.u-psud.fr}
\affiliation{Laboratoire de Physique Th\'eorique et Mod\`eles Statistiques, Universit\'e Paris-Sud, B\^at 100, 91405 Orsay Cedex, France.}
\date{\today}
\begin{abstract}
In the context of order statistics of discrete time random walks (RW), we investigate the statistics of the gap, $G_n$, and the number of time steps, $L_n$, between the two highest positions of a Markovian one-dimensional random walker, starting from $x_0 = 0$, after $n$ time steps (taking the $x$-axis vertical). The jumps $\eta_i = x_i - x_{i-1}$ are independent and identically distributed random variables drawn from a symmetric probability distribution function (PDF), $f(\eta)$, the Fourier transform of which has the small $k$ behavior $1 - \hat f(k) \propto |k|^\mu$, with $0 < \mu \leq 2$. For $\mu=2$, the variance of the jump distribution is finite and the RW (properly scaled) converges to a Brownian motion. For $0<\mu<2$, the RW is a L\'evy flight of index $\mu$. We show that the joint PDF of $G_n$ and $L_n$ converges to a well defined stationary bi-variate distribution $p(g,l)$ as the RW duration $n$ goes to infinity. We present a thorough analytical study of the limiting joint distribution $p(g,l)$, as well as of its associated marginals $p_{\rm gap}(g)$ and $p_{\rm time}(l)$, revealing a rich variety of behaviors depending on the tail of $f(\eta)$ (from slow decreasing algebraic tail to fast decreasing super-exponential tail). We also address the problem for a random bridge where the RW starts and ends at the origin after $n$ time steps. We show that in the large $n$ limit, the PDF of $G_n$ and $L_n$ converges to the {\it same} stationary distribution $p(g,l)$ as in the case of the free-end RW. Finally, we present a numerical check of our analytical predictions. Some of these results were announced in
a recent letter [S. N. Majumdar, Ph. Mounaix, G. Schehr, Phys. Rev. Lett. {\bf 111}, 070601 (2013)].
\end{abstract}
\maketitle
%
%
\section{Introduction and summary of main results}\label{sec1}

Random walks are not only cornerstones of statistical mechanics \cite{chandrasekhar,feller,hughes}, they have also found many applications in various areas of science, such as biology \cite{koshland}, computer science \cite{asmussen} or finance \cite{williams} to name just a few. Recently, they were shown to be a very good laboratory to study extreme and order statistics of strongly correlated random variables \cite{SM_book}. Indeed, although extreme value questions are very well understood in the case of independent and identically distributed (i.i.d.) random variables, much less is known in the case of strongly correlated variables. Random walks \cite{AS2005,SM2010,SM12,MMS,SOS_Airy} and Brownian motion \cite{Satya_Airy1,Satya_Airy2}, together with some rare examples like random matrix theory \cite{TW}, are scarce physically relevant models where exact analytical solutions have been obtained.  

While extreme value statistics usually focuses on the global maximum, $x_{\max}$, among a set of $n+1$ random variables, $x_0, \cdots, x_n$, many situations turn out to be more sensitive to the ``crowding'' near the maximum \cite{SM07}. Such situations occur when events whose amplitude is close to the maximum, which we call ``near extreme events'', become important. This is the case, for instance, in statistical seismology \cite{Omo1894,Uts61} or in finance \cite{LM03,PWHS10} where near extreme events like aftershocks or foreshocks may have drastic effects. The notion of near extreme events is also relevant for the physics of complex and disordered systems where the finite low temperature properties are dominated by the low lying states \cite{BM97,DM01,LDM03}, i.e. states with an energy close to the one of the ground state. 

A natural way to quantify this phenomenon of ``crowding'' near an extreme value makes use of the so-called density of states (close to the maximum or the minimum). This quantity, which counts the number of events whose amplitude is at a given distance from the maximum (or the minimum), was first studied for i.i.d. random variables both in mathematical statistics \cite{PL98} and in physics \cite{SM07}, where it was shown to exhibit non trivial and interesting behaviors \cite{SM07}. More recently, it has been studied for two types of strongly correlated random variables: Brownian motion \cite{PCMS13} and random Hermitian matrices \cite{PS14}. Here, we focus on yet another way to characterize the statistics of near extreme events: the order statistics of a set of ordered random variables, $x_{\max} = M_{1,n} > M_{2,n} \cdots > M_{n+1,n} = x_{\min}$, and the associated statistics of the gaps $d_{k,n} = M_{k,n} - M_{k+1,n}$ between the $k$-th and the $(k+1)$-th maxima which are particularly sensitive to the crowding. Unlike the order statistics of i.i.d. random variables which is now well understood~\cite{order_book}, little has been done so far regarding strongly correlated random variables and it is only recently that a few exact results were obtained for branching Brownian motion \cite{derrida_bbm,ramola_bbm} and random walks \cite{racz_order,SM12,MMS}.

In this paper, we consider a random walk starting at the origin, $x_0 = 0$, and evolving according to
\begin{equation}\label{def_RW}
x_i = x_{i-1} + \eta_{i},
\end{equation}
where the jumps $\eta_i$ are i.i.d. random variables distributed following a symmetric, bounded and piecewise continuous distribution $f(\eta)$ the Fourier transform of which, $\hat f(k) = \int_{-\infty}^{+\infty} f(\eta) {\rm e}^{i k \eta} d \eta$, has the small $k$ behavior
\begin{equation}\label{def_mu}
\hat f(k) = 1 - |a k|^\mu + o(|k|^\mu),
\end{equation}
where $0 < \mu \leq 2$ is the L\'evy index and $a>0$ is the characteristic length scale of the jumps. For $\mu = 2$, the variance of the jump distribution $\sigma^2 = \int_{-\infty}^{+\infty} \eta^2 f(\eta) d\eta$ is well defined and $a = \sigma /\sqrt{2}$. In this case the RW, suitably scaled, converges to a Brownian motion as $n\rightarrow +\infty$. On the other hand, for $0 < \mu < 2$, $f(\eta)$ does not possess a well defined second moment because of its heavy tails, $f(\eta) \propto |\eta|^{-1-\mu}$ ($\eta \to \infty$), and the RW\ (\ref{def_RW}) is a L\'evy flight of index $\mu$.

The analytical study performed in Ref. \cite{SM12} provided a first step in the study of order statistics of RW (\ref{def_RW}) restricted to the case $\mu = 2$ (jumps with a finite variance). It was shown in this work that the statistics of the gaps $d_{k,n}$ for the time series generated by the position of the RW after $n$ time steps becomes stationary, i.e. independent of $n$, as $n \to \infty$, confirming the results of previous numerical simulations \cite{racz_order}. The main focus of that paper \cite{SM12} was then on the study of the $k$-th stationary gap $d_{k,\infty} = \lim_{n \to \infty} d_{k,n}$ in the limit of large $k$. In this limit it was shown that the typical fluctuations of $d_{k,\infty}$ are of order ${O}(k^{-1/2})$\ \cite{SM12} (see also \cite{racz_order} for numerical evidence of this property). Remarkably, the probability distribution function (PDF) $p_{{\rm gap},k}(g)$, defined by $p_{{\rm gap},k}(g) dg = {\rm Prob.}(d_{k,\infty} \in [g, g+dg])$, takes a scaling form $p_{{\rm gap},k}(g) \sim (\sqrt{k}/\sigma) {\cal F}(g \sqrt{k}/\sigma)$, where $\sigma$ is the variance of the jump distribution and ${\cal F}(x)$ is a universal scaling function with an unexpected power law tail, ${\cal F}(x) \propto x^{-4}$, for large $x$. This scaling function ${\cal F}(x)$ was computed explicitly for an exponential jump distribution (in which case the full order statistics is exactly solvable for any finite $n$) and argued to be independent of the jump distribution, provided $\mu = 2$ in (\ref{def_mu}), on the basis of numerical simulations~\cite{SM12}.

We recently continued this work beyond the case $\mu = 2$ and performed an analytical study of the statistics of the first gap $d_{1,n}= M_{1,n} - M_{2,n}$, from now on denoted by $G_n$, and of the time $L_n$ elapsed between the corresponding first two maxima, for any value of $0< \mu \leq 2$\  \cite{MMS}. The first gap is a natural observable in the context of disordered systems (albeit usually between the first two minima). It is also relevant for e.g. applications in seismology, as it can model the difference in magnitude between the main shock and its largest aftershock \cite{Bat65,Ver69,CLMR03}. The time interval $L_n$ is of interest in e.g. seismology \cite{mega_prl}, financial markets \cite{palatella}, and queuing theory~\cite{cobham,grinstein}. 
\begin{center}
\begin{figure}[hht]
\includegraphics[width=0.8\linewidth]{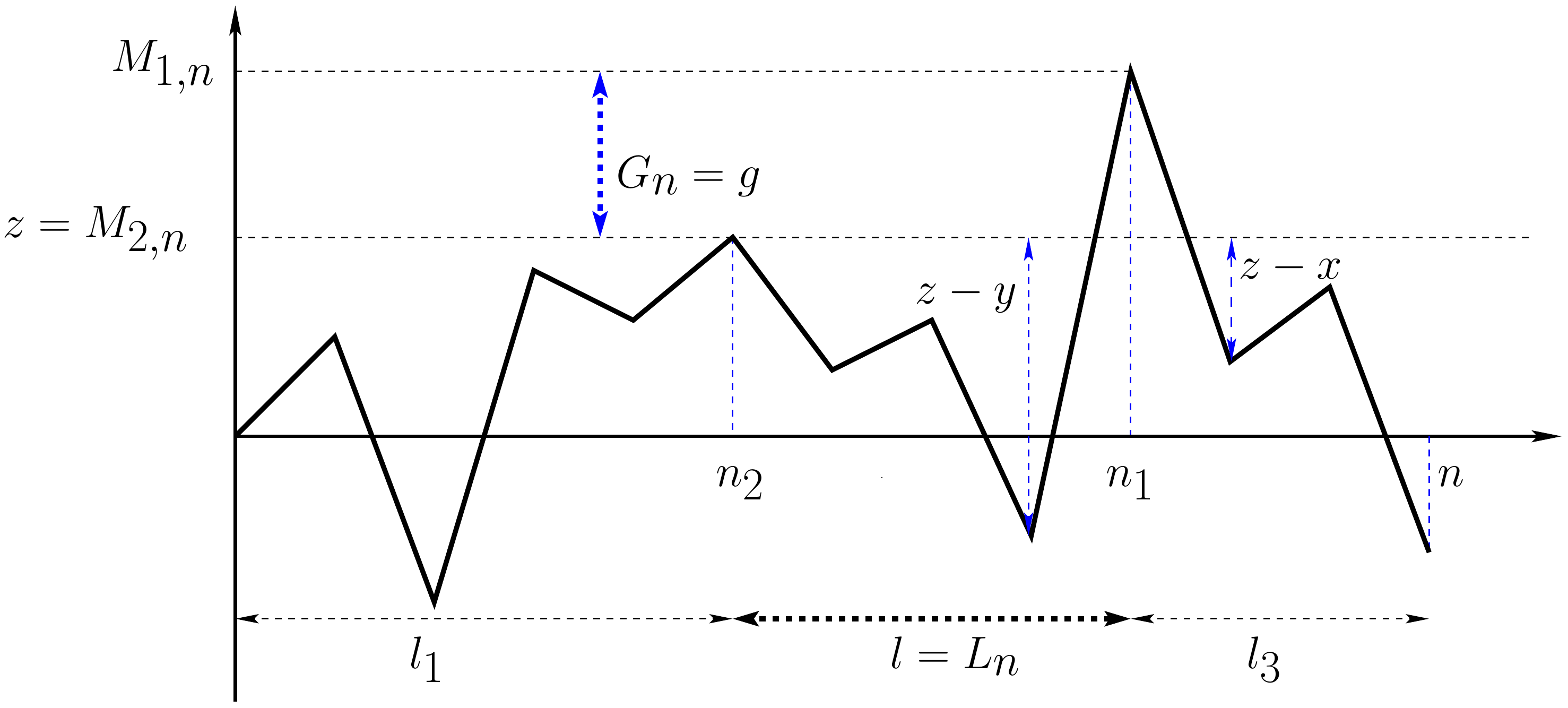}
\caption{Realization of a random walk (\ref{def_RW}) of $n$ steps. The indicated heights $x, y$ and $z$ correspond to the variables $x, y$ and $z$ in Eq. (\ref{sequence}). Here we focus on the joint PDF $p_n(g,l)$ of $G_n$ and $L_n$ in the limit of large $n$.}\label{Fig_markov}
\end{figure}
\end{center}

One goal of the present paper is to provide an exhaustive discussion of the different behaviors of the joint PDF of $G_n$ and $L_n$ that may arise in the large $n$ limit, depending on the large argument behavior of the jump PDF $f(\eta)$. In doing so we will give a detailed account of the results obtained in Ref. \cite{MMS}. We will complete our study by considering the case of a random bridge, i.e. the RW\ (\ref{def_RW}) conditioned to start and end at the origin, $x_n=x_0=0$. 

Before entering the details of the calculations, based to some extent on ideas and methods of first-passage problems \cite{Redner_book,Satya_review,Bray_review}, it is useful to summarize our main results. We first show that the joint PDF $p_n(g,l)$ of the random variables $G_n$ and $L_n$, with $G_n \in \mathbb{R}^{+}$ and $L_n \in \mathbb Z^*$, has a well defined limiting  PDF as $n \to \infty$:
\begin{equation}\label{limit_free}
\lim_{n \to \infty} p_n(g,l) = p(g,l),
\end{equation}
where the generating function (GF) of $p(g,l)$ with respect to (w.r.t.) $l$ is given in Eqs. (\ref{eq2.16}) and (\ref{eq2.17}). By summing the joint PDF $p(g,l)$ over $l \in {\mathbb Z}^*$ one obtains an exact expression for the marginal distribution $p_{\rm gap}(g)$, Eq.\ (\ref{eq3.1}), the full form of which depends on $f(\eta)$. For $\mu = 2$, there are some specific cases in which this full form can be computed explicitly, like e.g. gamma-distributed jumps for which $f(\eta) = \frac{b^{1+k}}{2k!}|\eta |^k \exp{(-b\,|\eta |)}$ with $k=0, 1$ (see Sec.\ \ref{sec3.1}). In such cases and more generally, the whole form of $p_{\rm gap}(g)$, including its tail, turns out to be quite sensitive to the jump distribution $f(\eta)$ (it can even have an algebraic tail if $f(\eta)$ also has a power law tail with finite $\sigma$). On the other hand, for $0 <\mu < 2$, we show that $p_{\rm gap}(g)$ has a generic algebraic tail
\begin{equation}\label{eq:result_p_of_g}
p_{\rm gap}(g) \underset{g \to \infty}{\sim} a^\mu {\tilde C_\mu} \; {g^{-1-\mu}},
\end{equation}
where ${\tilde C_\mu}$ is given in Eq.\ (\ref{eq3.22}). From this result (\ref{eq:result_p_of_g}) it follows immediately that the average gap diverges for $0<\mu\le 1$ while it is finite for $1<\mu <2$ (it is also finite for $\mu =2$). 

By integration of the joint PDF $p(g,l)$ over $g$, one finds that the marginal distribution $p_{\rm time}(l)$ displays a power law tail with an exponent depending only on $\mu$ as follows \footnote{Note that Ref. \cite{MMS} mentioned the existence of logarithmic corrections for $0<\mu <1$ and $1/\mu \in {\mathbb N}^*$: a careful analysis of these specific cases performed here in Appendix \ref{app3} shows instead that such logarithmic corrections are not present (except for $\mu = 1$ which plays here the role of a ``critical'' value).}:
 \begin{equation}\label{eq:result_p_of_l}
p_{\rm time}(l) \underset{l \to \infty}{\sim} 
\begin{cases}
{{\cal A}_{\rm I}}\,{l^{-1-1/\mu}} & \;, \; 1 < \mu \leq 2 \\
{{\cal A}_{\rm II} \, (\log l)}\,{l^{-2}} & \;, \, \mu = 1 \\
{{\cal A}_{\rm III}}\,{l^{-2}} & \;, \; 0 <\mu < 1 \;,
\end{cases}
\end{equation}
where the amplitudes ${\cal A}_{\rm I}, {\cal A}_{\rm II}$ and ${\cal A}_{\rm III}$ are given in Eq.\ (\ref{eq5.6}). Note that for $\mu=2$, one has $p_{\rm time}(l) \propto l^{-3/2}$ whatever the jump distribution $f(\eta)$ possessing a second moment. The third line of\ (\ref{eq:result_p_of_l}) reveals an unexpected freezing phenomenon of the exponent characterizing the algebraic tail of $p_{\rm time}(l)$ as $\mu$ decreases past the value $\mu_c = 1$. Interestingly, it follows from\ (\ref{eq:result_p_of_l}) that the first moment of $p_{\rm time}(l)$ is never defined. This means that, although the typical size of $L_n$ is ${O}(1)$, its average diverges with $n$. From (\ref{eq:result_p_of_l}) one can estimate that $\langle\vert L_n\vert\rangle \sim n^{1-1/\mu}$ for $1< \mu \le 2$, while $\langle\vert L_n\vert\rangle \sim \log{n}$ for $0 < \mu < 1$ and $\langle\vert L_n\vert\rangle\sim\ln(n)^2$ for $\mu =1$. 

These results follow from a detailed analysis of $p(g,l)$ in the plane $(g,l)$ for $\mu$ in the whole range $0 < \mu \leq 2$. We identify three different main types of behavior associated with (i) sub-exponential, (ii) exponential, and (iii) super-exponential decays of $f(\eta)$ at large $\eta$. L\'evy flights belong to type (i) behavior since, for $0 < \mu < 2$, $f(\eta)$ necessarily has an algebraic, sub-exponential, tail. In this case we show that in the scaling regime $g,\ l \gg 1$ with fixed $l g^{-\mu}$, the joint PDF $p(g,l)$ takes the following scaling form
\begin{equation}\label{eq:intro_joint}
p(g,l)\sim\frac{a^{2\mu}}{g^{1+2\mu}}F_\mu\left(\frac{a^\mu l}{g^\mu}\right)
 \ \ \ \ (l,g\rightarrow +\infty),
\end{equation}
with the asymptotic behaviors
\begin{equation}\label{joint_asympt}
F_\mu(y) \sim\left\lbrace
\begin{array}{ll}
B_\mu \, y^{-1/2}& y \to 0, \\
A_\mu  \, y^{-1-1/\mu}& y \to \infty ,
\end{array}
\right.
\end{equation}
where the amplitudes $B_\mu$ and $A_\mu$ are given in Eqs.\ (\ref{eq4.1.2.2}) and (\ref{eq4.1.2.5}), respectively. For $\mu = 2$, the three different types of behaviors mentioned above are possible, depending on the tail of $f(\eta)$ at large $\eta$. Interestingly, in the third case (iii) we find that the fast decrease of $f(\eta)$ gives rise to a concentration of $p(g,l)$ onto the two symmetric values $l = \pm 1$ as $g$ gets large, meaning that configurations with adjacent first and second maxima become much more likely in this limit. Such a concentration does not exist in the two other cases (i) and~(ii).

Finally, we finish our study by considering the case of a bridge, i.e. the RW\ (\ref{def_RW}) conditioned to start and end at the origin, $x_n=x_0=0$. We compute the joint PDF $p_n^{br}(g,l)$ of the first gap, $G_n$, and the time between the first two maxima, $L_n$, for the bridge after $n$ time steps. We show that $p_n^{br}(g,l)$ converges to a well-defined stationary distribution as $n\rightarrow +\infty$ which coincides exactly with the one for the free-end RW\ (\ref{def_RW}):
\begin{eqnarray}\label{limit_bridge}
\lim_{n \to \infty} p_n^{br}(g,l) = p(g,l) \;,  
\end{eqnarray}
where $p(g,l)$ is the same as in \ (\ref{limit_free}). It follows immediately that all the aforementioned results also hold for a random bridge without any change. 

The outline of the paper is as follows. In Section\ \ref{sec2} we derive the generating function of $p(g,l)$ with respect to $l$ for a free-end random walk. Section\ \ref{sec3} deals with the marginal distribution of the gap, $p_{\rm gap}(g)$. In Section\ \ref{sec3.1} we give the explicit expression of $p_{\rm gap}(g)$ for two particular cases (substantially, symmetrically gamma-distributed jumps with shape parameter equals to $1$ and $2$). Section\ \ref{sec3.2} gives the tail of $p_{\rm gap}(g)$ when the random walk is a L\'evy flight of index $\mu$ (with $0<\mu <2$). The asymptotic behavior of $p(g,l)$ for large $g$ or $l$ is investigated in Section\ \ref{sec4}. The two limits $l\rightarrow +\infty$ at fixed $g$ and $g\rightarrow +\infty$ at fixed $l$ are considered successively, the latter for three classes of jump distributions encompassing a wide range of jumps of practical interest (Sections\ \ref{sec4.1},\ \ref{sec4.2}, and\ \ref{sec4.3}). The behavior of $p(g,l)$ when both $g$ and $l$ are large is given and it is shown in Section\ \ref{sec4.1} that when the jump distribution has an algebraic tail, this behavior takes on a scaling form the derivation of which is given in Appendix\ \ref{app4}. Concentration of $p(g,l)$ onto $l=\pm 1$ as $g\rightarrow +\infty$ for jump distributions with a fast enough decreasing tail (e.g. super-exponential tail) is proved in Section\ \ref{sec4.3}. In Section\ \ref{sec5} we use the results obtained in Section\ \ref{sec4} to determine the tail of $p_{\rm time}(l)$ and show its freezing when the random walk is a L\'evy flight with index $0<\mu <1$. In Section\ \ref{sec6} we derive the generating function of $p(g,l)$ with respect to $l$ for a bridge and find that it is identical to the one obtained in Section\ \ref{sec2} for a free-end random walk. Finally, Section\ \ref{sec7} is devoted to the comparison of our analytical results with numerical simulations.
%
%
\section{Generating function of $\bm{p(g,l)}$ for a free-end random walk}\label{sec2}
Here we suppose without loss of generality that $n_1>n_2$. The starting point of our analysis is an exact formula for the joint PDF $p_n(g,l_1,l,l_3)$ of the gap $G_n=g$ and the three durations $l_1=n_2$, $l=n_1-n_2 = L_n$, and $l_3=n-n_1$ (see Fig. 1).

\begin{figure}
\includegraphics[width=\linewidth]{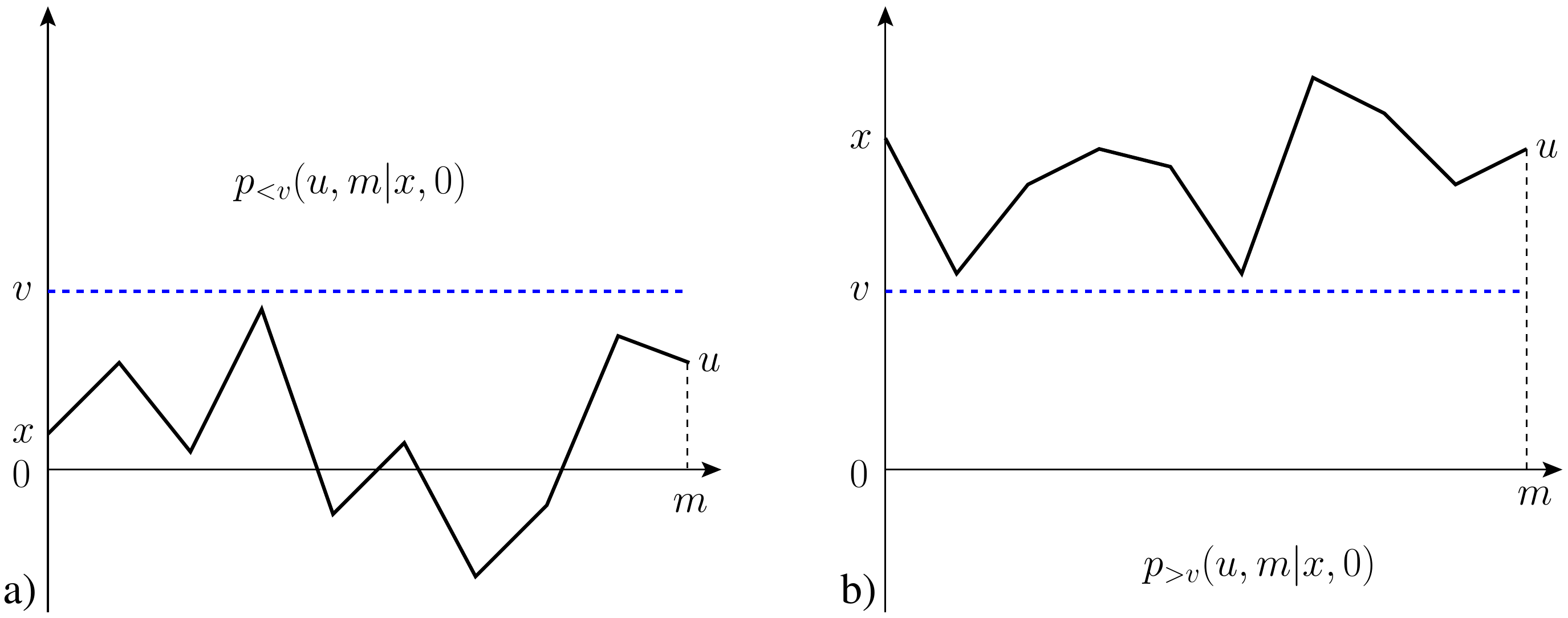}
\caption{a) A trajectory contributing to the Green's function $p_{<v}(u,m|x,0)$ defined in the text. b)~A trajectory contributing to the Green's function $p_{>v}(u,m|x,0)$ defined in the text.}\label{fig_propag}
\end{figure}

First, we set the notation which will be used in this section and in Sec.\ \ref{sec6} on the bridge. For given $u$ and $v$ on the real (vertical) line along which the random walker travels, we introduce:
\begin{itemize}
\item [$\bullet$] $p_{<v}(u,m\vert x,0)$ which is the Green's function (propagator) for a random walker starting at $x\le v$, arriving at $u\le v$ after $m$ steps and staying below $v$ in between (see Fig. \ref{fig_propag} a)). We denote by $P_{<v}(y<u,m\vert x,0) = \int_{-\infty}^u p_{<v}(y,m\vert x,0) \, dy$ the probability for the walker, starting at $x\le v$, to stay below $v$ up to step $m$ and to arrive anywhere in the interval $(-\infty ,u\rbrack$ at step~$m$.
\item [$\bullet$] $p_{>v}(u,m\vert x,0)$ which is the Green's function (propagator) for a random walker starting at $x\ge v$, arriving at $u\ge v$ after $m$ steps and staying above $v$ in between (see Fig. \ref{fig_propag} b)). We denote by $P_{>v}(y<u,m\vert x,0) = \int_{u}^\infty p_{>v}(y,m\vert x,0) \, dy$ the probability for the walker, starting at $x\ge v$, to stay above $v$ up to step $m$ and to arrive anywhere in the interval~$[u, +\infty)$ at step~$m$.
\end{itemize}

These Green's functions are important quantities to study the joint PDF $p_n(g,l_1,l,l_3)$ which we compute by considering the following sequence (see Fig. 1): (i) the walker starts from $0$ at time $0$ and reaches the second maximum $M_{2,n}=z$ at time $l_1$; (ii) then, starting from the second maximum it arrives at the position $x_{n_1-1}=y<z$ after $l-1$ steps, always staying below $z$, and makes a jump of amplitude $M_{n,1}-x_{n_1-1} = M_{n,2}+G_n-x_{n_1-1} = z+g-y$ to reach the first maximum; (iii) finally, if $n_1<n$ ($l_3>0$), it jumps from $M_{n,1}=M_{n,2}+G_n=z+g$ to $x_{n_1+1}=x<z$ and travels the last $l_3-1$ steps always staying below $z$. If $n_1=n$, i.e. the maximum is reached at the last step ($l_3=0$), the walker stops after stage (ii). Using the notation specified above, it is easily seen that such a sequence contributes with a probability
\begin{eqnarray}\label{sequence}
&&{\rm sequence\ contribution}=dx\, dy\, dz\, dg\, \times \\
&&\stackrel{\underbrace{p_{<z}(z,l_1\vert 0,0)}}{(i)}
\, \stackrel{\underbrace{p_{<z}(y,l-1\vert z,0)f(z+g-y)}}{(ii)}
\, \stackrel{\underbrace{f(z+g-x)P_{<z}(x_n <z,l_3-1\vert x,0)}}{(iii)} \nonumber
\end{eqnarray}
from which $p_n(g,l_1,l,l_3)$ is readily obtained by integrating over $x<z$, $y<z$, and $z>0$. One finds,
\begin{eqnarray}\label{eq2.1}
p_n(g,l_1,l,l_3)&=&\left\lbrack\int_0^{+\infty}p_{<z}(z,l_1\vert 0,0)
\left(\int_{y<z}p_{<z}(y,l-1\vert z,0)f(z+g-y)\, dy\right)\right. \nonumber \\
&\times&\left.\left(\int_{x<z}f(z+g-x)P_{<z}(x_n <z,l_3-1\vert x,0)\, dx\right)\, dz
\right\rbrack\, \delta_{l_1+l+l_3,n},
\end{eqnarray}
where the Kronecker delta function at the end of the right-hand side makes explicit the constraint $l_1+l+l_3=n$. In the second parenthesis we can write $P_{<z}(x_n <z,l_3-1\vert x,0)=P_{>0}(x_n >0,l_3-1\vert z-x,0)$, obtained by taking $z$ as a new origin and reversing space direction:
\begin{eqnarray}\label{eq2.2}
&&\int_{x<z}f(z+g-x)P_{<z}(x_n <z,l_3-1\vert x,0)\, dx \nonumber \\
&&= \int_{x<z} f(z+g-x)P_{>0}(x_n >0,l_3-1\vert z-x,0)\, dx= \nonumber \\
&&\int_{x>0}f(g+x)P_{>0}(x_n >0,l_3-1\vert x,0)\, dx= \nonumber \\
&&\int_0^{+\infty}q_{l_3-1}(x)f(g+x)\, dx = w_3(g,l_3),
\end{eqnarray}
independent of $z$, with 
\begin{eqnarray}\label{def_qm}
q_m(x)\equiv P_{>0}(x^\prime >0,m\vert x,0)\;,
\end{eqnarray}
and where, in the third line of Eq. (\ref{eq2.2}), we have made the change of variable $x\rightarrow z-x$. Similarly, we can write $p_{<z}(y,l-1\vert z,0)=p_{>0}(z-y,l-1\vert 0,0)$ in the first parenthesis:
\begin{eqnarray}\label{eq2.3}
&&\int_{y<z}p_{<z}(y,l-1\vert z,0)f(z+g-y)\, dy= \nonumber \\
&&\int_{y<z}p_{>0}(z-y,l-1\vert 0,0)f(z+g-y)\, dy= \nonumber \\
&&\int_{y>0}p_{>0}(y,l-1\vert 0,0)f(g+y)\, dy= \nonumber \\
&&\int_0^{+\infty}p_{l-1}(y)f(g+y)\, dy = w_2(g,l),
\end{eqnarray}
also independent of $z$, with 
\begin{eqnarray}\label{def_pm}
p_m(x)\equiv p_{>0}(x,m\vert 0,0) \;,
\end{eqnarray}
and where, in the third line of Eq. (\ref{eq2.3}), we have made the change of variable $y\rightarrow z-y$. Finally, in the remaining integral over $z$ we write $p_{<z}(z,l_1\vert 0,0)=p_{>0}(z,l_1\vert 0,0)$, obtained by taking $z$ as a new origin and reversing the directions of both space and time:
\begin{equation}\label{eq2.4}
\int_0^{+\infty}p_{<z}(z,l_1\vert 0,0)\, dz=
\int_0^{+\infty}p_{>0}(z,l_1\vert 0,0)\, dz=q_{l_1}(0)=w_1(l_1).
\end{equation}
Injecting\ (\ref{eq2.2}),\ (\ref{eq2.3}), and\ (\ref{eq2.4}) into the right-hand side of\ (\ref{eq2.1}) yields
\begin{equation}\label{eq2.5}
p_n(g,l_1,l,l_3)=w_1(l_1)w_2(g,l)w_3(g,l_3)\, \delta_{l_1+l+l_3,n}.
\end{equation}
Note that the last line of\ (\ref{eq2.2}) defines $w_3(g,l_3)$ for $l_3>0$. For realizations with $n_1=n$, ($l_3=0$), the second parenthesis on the right-hand side of\ (\ref{eq2.1}) is absent and one must set $w_3(g,l_3=0)=1$ in Eq.\ (\ref{eq2.5}), which completes the definition of $w_3(g,l_3)$ for all $l_3\ge 0$.

Thus, the problem is entirely determined by the two objects $q_m(x)$ (\ref{def_qm}) and $p_m(x)$ (\ref{def_pm}) a complete characterization of which is given by the Laplace transform with respect to $x$ of their generating functions with respect to $m$. Using the Hopf-Ivanov formula\ \cite{Iv}, one finds
\begin{equation}\label{eq2.6}
\int_0^{+\infty}\sum_{m\ge 0}q_m(x)s^m\, {\rm e}^{-\lambda x}dx=
\frac{\phi(\lambda ,s)}{\lambda\sqrt{1-s}},
\end{equation}
the so-called Pollaczek-Spitzer formula\ \cite{Pol,Spi,SM2010}, and
\begin{equation}\label{eq2.7}
\int_0^{+\infty}\sum_{m\ge 0}p_m(x)s^m\, {\rm e}^{-\lambda x}dx=
\phi(\lambda ,s),
\end{equation}
where the function $\phi(\lambda ,s)$ is given by (see Appendix\ \ref{app1})
\begin{equation}\label{eq2.8}
\phi(\lambda ,s)=\exp\left(-\frac{\lambda}{\pi}
\int_0^{+\infty}\frac{\ln\lbrack 1-s\hat{f}(k)\rbrack}{k^2+\lambda^2}\, dk\right) .
\end{equation}
We will not elaborate here on the Hopf-Ivanov formula and how it leads to Eqs.\ (\ref{eq2.6}) and\ (\ref{eq2.7}). The interested reader will find more details in Appendix\ \ref{app1} (see also \cite{MCZ2006}). Note that the quantities $q_m(x)$ (\ref{def_qm}) and $p_m(x)$ (\ref{def_pm}) have also proved very useful to compute the statistics of records of random walks~\cite{wergen_records}.

The joint PDF $p_n(g,l)$ is obtained formally from\ (\ref{eq2.5}) as $p_n(g,l)=\sum_{l_1,l_3\ge 0}p_n(g,l_1,l,l_3)$. Using Eqs.\ (\ref{eq2.2})-(\ref{eq2.4}) together with\ (\ref{eq2.6}) and\ (\ref{eq2.7}), one obtains an explicit expression for the double generating function of $p_n(g,l)$ with respect to $n$ and $l$ [we recall that $p_n(g,l)=p_n(g,-l)$\ \cite{note1}]. Namely,
\begin{eqnarray}\label{eq2.9}
&&\sum_{l,n>0}p_n(g,l)s^lt^n=
\sum_{l,n>0}\sum_{l_1,l_3\ge 0}p_n(g,l_1,l,l_3)s^lt^n \nonumber \\
&&=\left(\sum_{l_1\ge 0}w_1(l_1)t^{l_1}\right)
\left(\sum_{l>0}w_2(g,l)(st)^l\right)
\left(\sum_{l_3\ge 0}w_3(g,l_3)t^{l_3}\right) \\
&&=\left(\sum_{l_1\ge 0}w_1(l_1)t^{l_1}\right)
\left(\sum_{l>0}w_2(g,l)(st)^l\right)
\left(1+\sum_{l_3>0}w_3(g,l_3)t^{l_3}\right) ,\nonumber
\end{eqnarray}
where we have isolated the term with $l_3=0$ from the sum in the third parenthesis. The sum over $l_1$ can be computed explicitly by using Eq.\ (\ref{eq2.4}) and either the $\lambda\rightarrow +\infty$ limit of the Pollaczek-Spitzer formula\ (\ref{eq2.6}) with $\phi(+\infty ,s)=1$ or the $\lambda\rightarrow 0$ limit of Eq.\ (\ref{eq2.7}) with $\phi(0 ,s)=1/\sqrt{1-s}$ and $q_m(0)=\int_0^{+\infty}p_m(x)\, dx$. One finds the celebrated Sparre Andersen formula\ \cite{SA53},
\begin{equation}\label{eq2.10}
\sum_{l_1\ge 0}w_1(l_1)t^{l_1} =\sum_{l_1\ge 0}q_{l_1}(0)t^{l_1}=\frac{1}{\sqrt{1-t}}.
\end{equation}
The two remaining sums can be written as integrals. Let $u(x,s)$ and $h(x,s)$ denote the inverse Laplace transforms of $\phi(\lambda ,s)$ and $\phi(\lambda ,s)/\lambda$, respectively:
\begin{equation}\label{eq2.11}
\begin{array}{l}
\int_0^{+\infty}u(x,s){\rm e}^{-\lambda x}dx=\phi(\lambda ,s), \\
\int_0^{+\infty}h(x,s){\rm e}^{-\lambda x}dx=\phi(\lambda ,s)/\lambda .
\end{array}
\end{equation}
[Note that by Eq.\ (\ref{eq2.7}), $u(x,s)=\sum_{m\ge 0}p_{m}(x)s^m$]. It follows from Eqs.\ (\ref{eq2.3}) and\ (\ref{eq2.7}) that
\begin{eqnarray}\label{eq2.12}
\sum_{l>0}w_2(g,l)(st)^l &=&
\int_0^{+\infty}\left(\sum_{l>0}p_{l-1}(y)(st)^l\right) f(g+y)\, dy \nonumber \\
&=&st \int_0^{+\infty}\left(\sum_{m\ge 0}p_{m}(y)(st)^m\right) f(g+y)\, dy \\
&=&st\int_0^{+\infty}u(y,st) f(g+y)\, dy, \nonumber
\end{eqnarray}
where $m=l-1$, and from Eqs.\ (\ref{eq2.2}) and\ (\ref{eq2.6}) one obtains
\begin{eqnarray}\label{eq2.13}
\sum_{l_3>0}w_3(g,l)t^{l_3} &=&
\int_0^{+\infty}\left(\sum_{l_3>0}q_{l_3-1}(x)t^{l_3}\right) f(g+x)\, dx \nonumber \\
&=&t \int_0^{+\infty}\left(\sum_{m\ge 0}q_{m}(x)t^m\right) f(g+x)\, dx \\
&=&\left(t/\sqrt{1-t}\right)\int_0^{+\infty}h(x,t) f(g+x)\, dx, \nonumber
\end{eqnarray}
where $m=l_3-1$. Thus, using Eqs.\ (\ref{eq2.10}),\ (\ref{eq2.12}), and\ (\ref{eq2.13}) in the last line of\ (\ref{eq2.9}), one obtains
\begin{eqnarray}\label{eq2.14}
\sum_{l,n>0}p_n(g,l)s^lt^n&=&\frac{st}{\sqrt{1-t}}\int_0^{+\infty}u(y,st) f(g+y)\, dy \nonumber \\
&\times&\left( 1+\frac{t}{\sqrt{1-t}}\int_0^{+\infty}h(x,t) f(g+x)\, dx\right) .
\end{eqnarray}

The large $n$ behavior of $p_n(g,l)$ is encoded in the large $n$ behavior of its generating function with respect to $l$, $\sum_{l>0}p_n(g,l)s^l$, which can be extracted from the behavior of\ (\ref{eq2.14}) in the vicinity of its dominant singularity as a function of $t$ by appropriate Tauberian theorems. For all $g>0$ and complex $s$ with $\vert s\vert <1$, it is readily seen that\ (\ref{eq2.14}) has a single dominant singularity  at $t=1$ near which it behaves like
\begin{eqnarray}\label{eq2.15}
\sum_{n>0}\tilde{p}_n(g,s)t^n &\sim&\frac{s}{1-t}\int_0^{+\infty}u(y,s) f(g+y)\, dy \nonumber \\
&\times&\int_0^{+\infty}h(x,1) f(g+x)\, dx,\ \ \ \ \ (t\rightarrow 1)
\end{eqnarray}
where we have written $\tilde{p}_n(g,s)=\sum_{l>0}p_n(g,l)s^l$, the generating function of $p_n(g,l)$ with respect to $l$. Now, it follows from Darboux's theorem\ \cite{Hen} and $t=1$ being a simple pole on the right-hand side of\ (\ref{eq2.15}) that $\tilde{p}(g,s)=\lim_{n\rightarrow +\infty}\tilde{p}_n(g,s)$ exists and is given by the residue
\begin{equation}\label{eq2.16}
\tilde{p}(g,s)=I_1(g,s)I_2(g),
\end{equation}
with
\begin{equation}\label{eq2.17}
\begin{array}{l}
I_1(g,s)=s\int_0^{+\infty}u(y,s) f(g+y)\, dy, \\
I_2(g)=\int_0^{+\infty}h(x,1) f(g+x)\, dx.
\end{array}
\end{equation}
Thus, $p_n(g,l)$ converges to a limiting distribution $p(g,l)$ as $n\rightarrow +\infty$ whose generating function with respect to $l$, $\tilde{p}(g,s)=\sum_{l>0}p(g,l)s^l$, is given by\ (\ref{eq2.16}). The existence of the limit $p(g,l)$ is a consequence of the strong correlations between the walker's positions. Such a limiting distribution (without a $n$-dependent rescaling) does not exist for i.i.d. random variables and more generally, as strongly suggested by the numerical results of  \cite{racz_order}, for weakly correlated random variables. Expression\ (\ref{eq2.16}) -- together with Eqs.\ (\ref{eq2.8}), \ (\ref{eq2.11}), and\  (\ref{eq2.17})-- constitutes the central result of our study from which the various behaviors of $p(g,l)$ can be derived.

As a conclusion to this section, it is interesting to notice that the subdominant term on the right-hand side of\ (\ref{eq2.14}), which behaves like $1/\sqrt{1-t}$ for $t\rightarrow 1$, gives a correction to $\tilde{p}_n(g,s)$ that goes to zero like $1/\sqrt{n}$ as $n\rightarrow +\infty$. This term is due to the walks with $n_1=n$ which, therefore, do not contribute significantly to $p_n(g,l)$ for $n$ large enough. This result suggests that $p_n(g,l)$ might well not be significantly affected by what happens near the end of the walk, like a boundary condition at the $n$th step, for $n$ large enough (and not at all for $n\rightarrow +\infty$). We will see in Sec.\ \ref{sec6} that it is actually so when a free-end walk is replaced with a bridge.
%
%
\section{The marginal distribution $\bm{p_{\rm gap}(g)}$}\label{sec3}
In this section, we focus on the marginal distribution of the gap, $p_{\rm gap}(g)$, which, for any jump distribution, is exactly given by
\begin{equation}\label{eq3.1}
p_{\rm gap}(g)=\sum_{\vert l\vert >0}p(g,l)=2\sum_{l>0}p(g,l)
=2\tilde{p}(g,1)=2I_1(g,1)I_2(g),
\end{equation}
where we have used $p(g,l)=p(g,-l)$ and the expression\ (\ref{eq2.16}) of $\tilde{p}(g,s)$ with $s=1$. For $\mu =2$ (jumps with a finite variance), there are some particular cases in which\ (\ref{eq3.1}) can be computed explicitly, as exemplified in the following subsection.
\subsection{Explicit expression for two particular cases ($\bm{\mu =2}$)}\label{sec3.1}
The first example corresponds to a symmetric exponential jump distribution, $f(\eta)=(b/2)\exp(-b\vert\eta\vert)$ with $b>0$. In this case, Eqs.\ (\ref{eq2.17}) and \ (\ref{eq2.11}) give
\begin{equation}\label{eq3.2}
I_1(g,1)=\frac{b}{2}{\rm e}^{-bg}\int_0^{+\infty}u(y,1){\rm e}^{-by}dy
=\frac{b}{2}\phi(b,1){\rm e}^{-bg},
\end{equation}
and
\begin{equation}\label{eq3.3}
I_2(g)=\frac{b}{2}{\rm e}^{-bg}\int_0^{+\infty}h(x,1){\rm e}^{-bx}dx
=\frac{1}{2}\phi(b,1){\rm e}^{-bg},
\end{equation}
and Eq.\ (\ref{eq3.1}) reduces to
\begin{equation}\label{eq3.4}
p_{\rm gap}(g)=\frac{b}{2}\phi(b,1)^2{\rm e}^{-2bg}.
\end{equation}
It remains to check that $\phi(b,1)=2$, as demanded by the normalization of $p_{\rm gap}(g)$. From\ (\ref{eq2.8}) with $\hat{f}(k)=\lbrack 1+(k/b)^2\rbrack^{-1}$, $s=1$, and $\lambda =b$, one finds
\begin{equation}\label{eq3.5}
\phi(b,1)=\exp\left\lbrack -\frac{1}{\pi}
\int_0^{+\infty}\frac{1}{1+q^2}\ln\left(\frac{q^2}{1+q^2}\right)\, dq\right\rbrack ,
\end{equation}
where $q=k/b$. The integral over $q$ is equal to $-\pi\ln 2$ and $\phi(b,1)=2$, as it should be. Thus, for a symmetric exponential jump distribution, $p_{\rm gap}(g)$ is exactly given by
\begin{equation}\label{eq3.6}
p_{\rm gap}(g)=2b{\rm e}^{-2bg}.
\end{equation}

In the second example we take $f(\eta)=(b^2/2)\vert\eta\vert\exp(-b\vert\eta\vert)$ with $b>0$. In this case, Eqs.~(\ref{eq2.11}) and (\ref{eq2.17}) give
\begin{eqnarray}\label{eq3.7}
I_1(g,1)&=&\frac{b^2}{2}{\rm e}^{-bg}
\left\lbrack g\int_0^{+\infty}u(y,1){\rm e}^{-by}dy+
\int_0^{+\infty}y\, u(y,1){\rm e}^{-by}dy\right\rbrack \nonumber \\
&=&\frac{b^2}{2}{\rm e}^{-bg}\left\lbrack g\phi(b,1)-\left.\frac{d\phi(\lambda ,1)}
{d\lambda}\right\vert_{\lambda =b}\right\rbrack ,
\end{eqnarray}
and
\begin{eqnarray}\label{eq3.8}
I_2(g)&=&\frac{b^2}{2}{\rm e}^{-bg}
\left\lbrack g\int_0^{+\infty}h(x,1){\rm e}^{-bx}dx+
\int_0^{+\infty}x\, u(x,1){\rm e}^{-bx}dx\right\rbrack \nonumber \\
&=&\frac{b^2}{2}{\rm e}^{-bg}\left\lbrack\frac{g}{b}\phi(b,1)-\left.\frac{d\lambda^{-1}\phi(\lambda ,1)}
{d\lambda}\right\vert_{\lambda =b}\right\rbrack \\
&=&\frac{1}{2}{\rm e}^{-bg}
\left\lbrack (bg+1)\phi(b,1)-b\left.\frac{d\phi(\lambda ,1)}
{d\lambda}\right\vert_{\lambda =b}\right\rbrack , \nonumber
\end{eqnarray}
which, together with Eq.\ (\ref{eq3.1}), lead to
\begin{eqnarray}\label{eq3.9}
p_{\rm gap}(g)&=&\frac{b}{2}{\rm e}^{-2bg}\left\lbrack bg\phi(b,1)-b\left.\frac{d\phi(\lambda ,1)}
{d\lambda}\right\vert_{\lambda =b}\right\rbrack \nonumber \\
&\times&\left\lbrack (bg+1)\phi(b,1)-b\left.\frac{d\phi(\lambda ,1)}
{d\lambda}\right\vert_{\lambda =b}\right\rbrack .
\end{eqnarray}
From\ (\ref{eq2.8}) with $\hat{f}(k)=\lbrack 1-(k/b)^2\rbrack/\lbrack 1+(k/b)^2\rbrack^2$, $s=1$, and $\lambda =b$, one finds
\begin{equation}\label{eq3.10}
\phi(b,1)=\exp\left\lbrack -\frac{1}{\pi}
\int_0^{+\infty}\frac{1}{1+q^2}\ln\left(\frac{q^2(3+q^2)}{(1+q^2)^2}\right)\, dq\right\rbrack ,
\end{equation}
and
\begin{equation}\label{eq3.11}
\left.\frac{d\phi(\lambda ,1)}{d\lambda}\right\vert_{\lambda =b}=
\frac{\phi(b,1)}{\pi b}\int_0^{+\infty}\frac{1-q^2}{(1+q^2)^2}
\ln\left(\frac{q^2(3+q^2)}{(1+q^2)^2}\right)\, dq\ ,
\end{equation}
where $q=k/b$. The integrals over $q$ in\ (\ref{eq3.10}) and\ (\ref{eq3.11}) are respectively equal to $\pi\ln\lbrack (1+\sqrt{3})/4\rbrack$ and $(1-\sqrt{3})\pi /2$. Using these results in Eq.\ (\ref{eq3.9}) one obtains, after some straightforward algebra,
\begin{equation}\label{eq3.12}
p_{\rm gap}(g)=c\lbrack (2bg+\sqrt{3})^2-1\rbrack {\rm e}^{-2bg},
\end{equation}
with $c=2b/(1+\sqrt{3})^2$.

From these two examples it can be seen that when the jumps of the walker have a finite variance ($\mu =2$), even the tail of $p_{\rm gap}(g)$ for large $g$ depends on the details of the jump distribution $f(\eta)$. Actually, as we will show in Sec.\ \ref{sec4}, it depends on the tail of $f(\eta)$ which, for $\mu =2$, can be very different from one case to the other. On the other hand, for $0<\mu <2$, both the tails of $f(\eta)$ and $p_{\rm gap}(g)$ depend on the L\'evy index $\mu$ only, as we will now see.
\subsection{Large $\bm{g}$ behavior for $\bm{0<\mu <2}$}\label{sec3.2}
For $0<\mu <2$, it turns out that the large $g$ limit of the integrals over $x$ and $y$ in Eq.~(\ref{eq2.17}) with $s=1$ are dominated by large values of $x,\ y\sim O(g)$. Hence, to study the large $g$ behavior of $p_{\rm gap}(g)$, one needs the large argument behavior of $u(y,1)$ and $h(x,1)$ which can in turn be obtained from the small $\lambda$ limit of $\phi(\lambda ,1)$. From Eq.\ (\ref{eq2.8}) in which one makes the change of variable $k=\lambda q$ and let $\lambda\rightarrow 0$\ \cite{note2}, one finds
\begin{eqnarray}\label{eq3.13}
\phi(\lambda ,1)&=&\exp\left(-\frac{1}{\pi}
\int_0^{+\infty}\frac{\ln\lbrack 1-\hat{f}(\lambda q)\rbrack}{1+q^2}\, dq\right) \nonumber \\
&\sim&\exp\left(-\frac{1}{\pi}
\int_0^{+\infty}\frac{\ln(a^\mu\lambda^\mu q^\mu)}{1+q^2}\, dq\right)
\ \ \ \ (\lambda\rightarrow 0) \\
&=&\exp\left\lbrack -\frac{\mu}{\pi}\left(\ln(a\lambda)\int_0^{+\infty}\frac{dq}{1+q^2}+
\int_0^{+\infty}\frac{\ln q}{1+q^2}\, dq\right)\right\rbrack
=\frac{1}{(a\lambda)^{\mu/2}}, \nonumber
\end{eqnarray}
where the first and second integrals over $q$ in the last line of\ (\ref{eq3.13}) are respectively equal to $\pi /2$ and $0$. Laplace inverting the first Eq.\ (\ref{eq2.11}) gives
\begin{equation}\label{eq3.14}
u(x,1)=\frac{1}{2i\pi}\int_{\cal{L}}\phi(\lambda ,1){\rm e}^{\lambda x}d\lambda =
\frac{1}{2i\pi x}\int_{\cal{L}}\phi(\overline{\lambda}/x ,1)
{\rm e}^{\overline{\lambda}}d\overline{\lambda} ,
\end{equation}
where $\overline{\lambda} =\lambda x$ and $\cal{L}$ is a Bromwich contour. Letting $x\rightarrow +\infty$ in\ (\ref{eq3.14}) and using\ (\ref{eq3.13}), one obtains
\begin{equation}\label{eq3.15}
u(x,1)\sim\frac{x^{\mu/2-1}}{a^{\mu/2}}\, \frac{1}{2i\pi}\int_{\cal{L}}\frac{\exp(\overline{\lambda})}
{\overline{\lambda}^{\mu/2}}\, d\overline{\lambda} = \frac{x^{\mu/2-1}}{a^{\mu/2}\Gamma(\mu/2)}
\ \ \ \ (x\rightarrow +\infty).
\end{equation}
Similarly, the second Eq.\ (\ref{eq2.11}) gives
\begin{equation}\label{eq3.16}
h(x,1)=\frac{1}{2i\pi}\int_{\cal{L}}\phi(\lambda ,1){\rm e}^{\lambda x}\, \frac{d\lambda}{\lambda} =
\frac{1}{2i\pi}\int_{\cal{L}}\phi(\overline{\lambda}/x ,1)
{\rm e}^{\overline{\lambda}}\, \frac{d\overline{\lambda}}{\overline{\lambda}} ,
\end{equation}
and
\begin{equation}\label{eq3.17}
h(x,1)\sim\frac{x^{\mu/2}}{a^{\mu/2}}\, \frac{1}{2i\pi}\int_{\cal{L}}\frac{\exp(\overline{\lambda})}
{\overline{\lambda}^{\mu/2+1}}\, d\overline{\lambda} = \frac{x^{\mu/2}}{a^{\mu/2}\Gamma(\mu/2+1)}
\ \ \ \ (x\rightarrow +\infty).
\end{equation}
[Note that Eq.\ (\ref{eq3.13}) and the resulting Eqs.\ (\ref{eq3.15}) and\ (\ref{eq3.17}) hold for $\mu =2$ too]. For $0<\mu <2$, the large $\eta$ behavior of $f(\eta)$ can be obtained from the small $k$ behavior of its Fourier transform $\hat{f}(k)$ (see Appendix\ \ref{app2}). One finds
\begin{equation}\label{eq3.18}
f(\eta)\sim \sin\left(\frac{\mu\pi}{2}\right)\Gamma(\mu+1)
\frac{a^\mu}{\pi \eta^{\mu+1}} \ \ \ \ (\eta\rightarrow +\infty).
\end{equation}
We now have all the ingredients we need to proceed. To determine the large $g$ behavior of $I_1(g,1)$ we make the change of variable $y=g\overline{y}$ in the first Eq.\ (\ref{eq2.17}) with $s=1$, let $g\rightarrow +\infty$, and use Eqs.\ (\ref{eq3.15}) and\ (\ref{eq3.18}). One obtains
\begin{eqnarray}\label{eq3.19}
I_1(g,1)&=&g\int_0^{+\infty}u(g\overline{y},1)
f\lbrack g(1+\overline{y})\rbrack\, d\overline{y} \nonumber \\
&\sim&\frac{a^{\mu/2}}{\pi}\sin\left(\frac{\mu\pi}{2}\right)
\frac{\Gamma(\mu+1)}{\Gamma(\mu/2)}\, \frac{1}{g^{1+\mu/2}}\,
\int_0^{+\infty}\frac{\overline{y}^{\mu/2-1}}{(1+\overline{y})^{\mu+1}}\, d\overline{y}
 \ \ \ \ (g\rightarrow +\infty) \nonumber \\
 &=&\frac{\mu a^{\mu/2}}{2\Gamma(1-\mu/2)}\, \frac{1}{g^{1+\mu/2}},
\end{eqnarray}
where the integral over $\overline{y}$ in the second line is equal to $\Gamma(\mu/2+1)\Gamma(\mu/2)/\Gamma(\mu+1)$ and we have used the reflection formula $\Gamma(z)\Gamma(1-z)=\pi/\sin(\pi z)$. Following the same line, from the second Eq.\ (\ref{eq2.17}) one gets
\begin{eqnarray}\label{eq3.20}
I_2(g)&=&g\int_0^{+\infty}h(g\overline{x},1)
f\lbrack g(1+\overline{x})\rbrack\, d\overline{x} \nonumber \\
&\sim&\frac{a^{\mu/2}}{\pi}\sin\left(\frac{\mu\pi}{2}\right)
\frac{\Gamma(\mu+1)}{\Gamma(\mu/2+1)}\, \frac{1}{g^{\mu/2}}\,
\int_0^{+\infty}\frac{\overline{x}^{\mu/2}}{(1+\overline{x})^{\mu+1}}\, d\overline{x}
 \ \ \ \ (g\rightarrow +\infty) \nonumber \\
 &=&\frac{a^{\mu/2}}{\Gamma(1-\mu/2)}\, \frac{1}{g^{\mu/2}}.
\end{eqnarray}
Note that the integral over $\overline{x}$ in the second line of\ (\ref{eq3.20}) is equal to the integral over $\overline{y}$ in the second line of\ (\ref{eq3.19}) (make the change of variable $\overline{y}=1/\overline{x}$). The large $g$ behavior of $p_{\rm gap}(g)$ for $0<\mu <2$ is obtained from Eqs.\ (\ref{eq3.1}),\ (\ref{eq3.19}), and\ (\ref{eq3.20}). One finds the algebraic tail
\begin{equation}\label{eq3.21}
p_{\rm gap}(g)\sim\frac{\tilde{C}_\mu a^\mu}{g^{1+\mu}} \ \ \ \ (g\rightarrow +\infty),
\end{equation}
with
\begin{equation}\label{eq3.22}
\tilde{C}_\mu =\frac{\mu}{\Gamma(1-\mu/2)^2}.
\end{equation}
From this result it follows immediately that the average gap diverges for $0<\mu\le 1$ while it is finite for $1<\mu <2$. (It is also finite for $\mu =2$).

The asymptotic behavior\ (\ref{eq3.21}) can be generalized to any distribution with an algebraic tail of the form $f(\eta)\sim C_\alpha d^\alpha /\eta^{\alpha +1}$, with $0<\alpha\ne 2$, $\mu =\min(2,\alpha)$, and where $C_\alpha$ is a constant, ($d$ a length scale not necessarily equal to $a$). After some straightforward algebra, one obtains
\begin{equation}\label{eq3.23}
p_{\rm gap}(g)\sim\frac{\tilde{C}_\alpha d^{2\alpha}}{a^\mu g^{1+2\alpha -\mu}}
 \ \ \ \ (g\rightarrow +\infty),
\end{equation}
with
\begin{equation}\label{eq3.24}
\tilde{C}_\alpha =2C_\alpha^2
\frac{\Gamma(1+\alpha -\mu/2)\Gamma(\alpha -\mu/2)}{\Gamma(\alpha +1)^2}.
\end{equation}
For $0<\alpha <2$, ($\alpha =\mu$), the algebraic tail is necessarily given by\ (\ref{eq3.18}) (see Appendix\ \ref{app2}) and it can be checked that\ (\ref{eq3.23}) and\ (\ref{eq3.24}) reduce respectively to\ (\ref{eq3.21}) and\ (\ref{eq3.22}), as it should be. We will get back to jump distributions with an algebraic tail in greater detail in Sec.\ \ref{sec4.1.1}.
%
%
\section{Asymptotic behavior of $\bm{p(g,l)}$ for large $\bm{g}$ or $\bm{l}$}\label{sec4}
In this section, which is the main course of the paper, we determine the asymptotic behavior of the joint PDF $p(g,l)$ when either $g$ or $l$ (or both) is large. From the results obtained for the large $l$ behavior of $p(g,l)$ we will then be able to derive the large $l$ behavior of the marginal distribution $p_{\rm{time}}(l)=\int_0^{+\infty}p(g,l)\, dg$, which will be the subject of the next section.

First, we consider the limit $l\rightarrow +\infty$ at fixed $g$. The behavior of $p(g,l)$ in this limit can be deduced from the generating function $\tilde{p}(g,s)$ near its dominant singularity at $s=1$, which in turn is entirely determined by $\phi(\lambda ,s)$ in the same vicinity of $s=1$ through\ (\ref{eq2.16}),\ (\ref{eq2.17}), and\ (\ref{eq2.11}). Let us rewrite\ (\ref{eq2.8}) as
\begin{equation}\label{eq4.1}
\phi(\lambda ,s)=\phi(\lambda ,1)\exp\left(-\frac{\lambda}{\pi}
\int_0^{+\infty}\frac{\ln\lbrack 1+(1-s)\hat{F}(k)\rbrack}{k^2+\lambda^2}\, dk\right) ,
\end{equation}
where
\begin{equation}\label{eq4.2}
\hat{F}(k)=\frac{\hat{f}(k)}{1-\hat{f}(k)}.
\end{equation}
To examine how\ (\ref{eq4.1}) behaves near $s=1$, we need to deal with integral and non-integral $1/\mu$ separately. Here, we take $0<\mu\le 2$ and non-integral $1/\mu$. (The case where $1/\mu$ is an integer is dealt with in Appendix\ \ref{app3}). Expanding the logarithm on the right-hand side of\ (\ref{eq4.1}) in power series of $(1-s)$ up to order $\lbrack 1/\mu\rbrack$, the integer part of $1/\mu$, we write
\begin{equation}\label{eq4.3}
\int_0^{+\infty}\frac{\lambda\, \ln\lbrack 1+(1-s)\hat{F}(k)\rbrack}{k^2+\lambda^2}\, dk
= \sum_{n=1}^{\lbrack 1/\mu\rbrack}\frac{(-1)^{n+1}}{n}(1-s)^n
\int_0^{+\infty}\frac{\lambda\, \hat{F}(k)^n}{k^2+\lambda^2}\, dk +R(\lambda ,s),
\end{equation}
where $R(\lambda ,s)$ is the remainder,
\begin{equation}\label{eq4.4}
R(\lambda ,s)=\int_0^{+\infty}\frac{\lambda}{k^2+\lambda^2}\left\lbrace
\ln\lbrack 1+(1-s)\hat{F}(k)\rbrack +
\sum_{n=1}^{\lbrack 1/\mu\rbrack}\frac{(-1)^n}{n}(1-s)^n\hat{F}(k)^n\right\rbrace\, dk.
\end{equation}
Since $\hat{F}(k)\sim k^{-\mu}$ as $k\rightarrow 0$, the integrals over $k$ in the sums over $n\le \lbrack 1/\mu\rbrack$ in\ (\ref{eq4.3}) and\ (\ref{eq4.4}) exist. To determine the asymptotic behavior of $R(\lambda ,s)$ near $s=1$, we make the change of variable $k=(1-s)^{1/\mu}q/a$ in\ (\ref{eq4.4}) and let $s\rightarrow 1$\ \cite{note3}. One gets
\begin{eqnarray}\label{eq4.5}
R(\lambda ,s)&\sim&\frac{(1-s)^{1/\mu}}{a\lambda}
\int_0^{+\infty}\left\lbrack\ln\left(1+\frac{1}{q^\mu}\right)+\sum_{n=1}^{\lbrack 1/\mu\rbrack}
\frac{(-1)^n}{nq^{n\mu}}\right\rbrack\, dq \ \ \ \ (s\rightarrow 1) \nonumber \\
&=&\frac{(1-s)^{1/\mu}}{a\mu\lambda}
\int_0^{+\infty}\left\lbrack\ln\left(1+u\right)+\sum_{n=1}^{\lbrack 1/\mu\rbrack}
\frac{(-u)^n}{n}\right\rbrack\, \frac{du}{u^{1+1/\mu}} \\
&=&\frac{\pi (1-s)^{1/\mu}}{a\lambda\sin(\pi/\mu)}, \nonumber
\end{eqnarray}
where the integral over $u=1/q^\mu$ in the second line is equal to $\mu\pi/\sin(\pi/\mu)$. Thus, the asymptotic expansion of\ (\ref{eq4.3}) near $s=1$ reads
\begin{equation}\label{eq4.6}
\int_0^{+\infty}\frac{\lambda\, \ln\lbrack 1+(1-s)\hat{F}(k)\rbrack}{k^2+\lambda^2}\, dk
=\sum_{n=1}^{\lbrack 1/\mu\rbrack}\frac{(-1)^{n+1}}{n}\beta_n(\lambda)(1-s)^n
+\pi a_\mu\frac{(1-s)^{1/\mu}}{\lambda}+o(1-s)^{1/\mu} ,
\end{equation}
with $a_\mu =1/\lbrack a\sin(\pi/\mu)\rbrack$ and
\begin{equation}\label{eq4.7}
\beta_n(\lambda)=\int_0^{+\infty}\frac{\lambda\, \hat{F}(k)^n}{k^2+\lambda^2}\, dk.
\end{equation}
From\ (\ref{eq4.1}) and\ (\ref{eq4.6}) one obtains the behavior of $\phi(\lambda ,s)$ in the vicinity of $s=1$ as
\begin{equation}\label{eq4.8}
\phi(\lambda ,s)=\phi(\lambda ,1)\left\lbrack 1+
\sum_{n=1}^{\lbrack 1/\mu\rbrack}\gamma_n(\lambda)(1-s)^n
-a_\mu\frac{(1-s)^{1/\mu}}{\lambda}+o(1-s)^{1/\mu}\right\rbrack,
\end{equation}
where the $\gamma_n(\lambda)$s are sums of products of the form $\prod_i\beta_{n_i}(\lambda)$ with $\sum_i n_i=n$. Eqs.\ (\ref{eq4.8}) and\ (\ref{eq2.11}) yield
\begin{equation}\label{eq4.9}
u(x,s)=u(x,1)+\sum_{n=1}^{\lbrack 1/\mu\rbrack}w_n(x)(1-s)^n
-a_\mu h(x,1)(1-s)^{1/\mu}+o(1-s)^{1/\mu},
\end{equation}
near $s=1$, with
\begin{equation}\label{eq4.10}
w_n(x)=\frac{1}{2i\pi}\int_{\cal{L}}\phi(\lambda ,1)\gamma_n(\lambda){\rm e}^{\lambda x}d\lambda .
\end{equation}
Injecting the expansion\ (\ref{eq4.9}) into the expression\ (\ref{eq2.17}) for $I_1(g,s)$ and putting the result in Eq.\ (\ref{eq2.16}), one obtains
\begin{equation}\label{eq4.11}
\tilde{p}(g,s)=\tilde{p}(g,1) +I_2(g)\sum_{n=1}^{\lbrack 1/\mu\rbrack}J_n(g)(1-s)^n
-a_\mu I_2(g)^2 (1-s)^{1/\mu}+o(1-s)^{1/\mu},
\end{equation}
near $s=1$, with
\begin{equation}\label{eq4.12}
J_n(g)=\int_0^{+\infty}w_n(x)\, f(g+x)\, dx.
\end{equation}
Since $n\le\lbrack 1/\mu\rbrack$ and $1/\mu$ is not an integer, the existence of the integral\ (\ref{eq4.12}) defining $J_n(g)$ is ensured by $w_n(x)\sim x^{\mu(n+1/2)-1}$ as $x\rightarrow +\infty$, which is readily obtained from Eq.\ (\ref{eq4.7}) and $\hat{F}(k)\sim k^{-\mu}$ as $k\rightarrow 0$ giving $\beta_n(\lambda)\sim 1/\lambda^{n\mu}$, hence $\gamma_n(\lambda)\sim 1/\lambda^{n\mu}$, as $\lambda\rightarrow 0$, and from $\phi(\lambda ,1)\sim 1/\lambda^{\mu/2}$ in the same limit [see Eq.\ (\ref{eq3.13})].

The large $l$ behavior of $p(g,l)$ at fixed $g$ is determined by the singular term $(1-s)^{1/\mu}$ on the right-hand side of\ (\ref{eq4.11}) through the appropriate Tauberian theorem (here Darboux's theorem\ \cite{Hen}, see also e.g. Theorem VI.1 in \cite{FS}). It turns out that a factor $(1-s)^{1/\mu}$,\ ($s\rightarrow 1$), in $\tilde{p}(g,s)$ translates into a factor $1/\lbrack\Gamma(-1/\mu)l^{1+1/\mu}\rbrack$,\ ($l\rightarrow +\infty$), in $p(g,l)$ and one gets, after some straightforward algebra,
\begin{equation}\label{eq4.13}
p(g,l)\sim\frac{\Gamma(1+1/\mu)}{\pi a}\, \frac{I_2(g)^2}{l^{1+1/\mu}} \ \ \ \ (l\rightarrow +\infty),
\end{equation}
where we have used $a_\mu =1/\lbrack a\sin(\pi/\mu)\rbrack$ and the reflection formula $\Gamma(-z)\Gamma(z+1)=-\pi/\sin(\pi z)$. [Note that $I_2(g)$ depends also on $a$].

The case where $1/\mu$ is an integer is dealt with in Appendix\ \ref{app3}. The algebraic singularity on the right-hand side of Eq.\ (\ref{eq4.8}) changes to a logarithmic singularity, but the asymptotic behavior of $p(g,l)$ for $l\rightarrow +\infty$ at fixed $g$ is again found to be given by Eq.\ (\ref{eq4.13}) which, therefore, holds in the whole interval $0<\mu\le 2$ without any restriction. (See Appendix\ \ref{app3} for details).

Now, we consider the limit $g\rightarrow +\infty$ at fixed $l$. If the support of the jump distribution $f(\eta)$ is bounded, the first gap cannot be larger than the (finite) diameter of this support and one trivially has $p(g,l)=0$ for every $l$ and $g>g_0$ with $g_0<+\infty$ large enough. The situation is more interesting if the support of $f(\eta)$ is not bounded. In this case, we will see in the following that the asymptotic behavior of $p(g,l)$ in the large $g$ limit depends on the tail of $f(\eta)$. We have singled out three classes of tails which encompass a wide range of jumps of practical interest.
%
%
\subsection{Class A jumps: slow decreasing $\bm{f(\eta)}$}\label{sec4.1}
Class A jumps are defined by
\begin{equation}\label{eq4.1.1}
f(g+x)\sim f(g) \ \ \ \ (g\rightarrow +\infty),
\end{equation}
for any fixed $x$. Jump distributions with an algebraic tail are class A. We will make a comprehensive study of this important particular case below. First we give some general results. The large $g$ behavior of $I_1(g,s)$ is readily obtained from\ (\ref{eq2.11}), (\ref{eq2.17}) and (\ref{eq4.1.1}). One gets
\begin{equation}\label{eq4.1.2}
I_1(g,s)\sim sf(g)\int_0^{+\infty}u(y,s)\, dy=sf(g)\phi(0,s)
=\frac{sf(g)}{\sqrt{1-s}} \ \ \ \ (g\rightarrow +\infty),
\end{equation}
where we have used $\phi(0,s)=1/\sqrt{1-s}$ as given by Eq.\ (\ref{eq2.8}) in the limit $\lambda\rightarrow 0$, (see also e.g.\ \cite{AS2005}). The large $g$ behavior of $I_2(g)$ depends on $f(\eta)$ and cannot be specified further on at this point. Thus, one has
\begin{equation}\label{eq4.1.3}
\tilde{p}(g,s)\sim\frac{sf(g)I_2(g)}{\sqrt{1-s}} \ \ \ \ (g\rightarrow +\infty),
\end{equation}
from which the large $g$ behavior of $p(g,l)$ at fixed $l$ for class A jumps is straightforwardly found to be given by
\begin{equation}\label{eq4.1.4}
p(g,l)\sim f(g)I_2(g)
\, \frac{\left(_{\vert l\vert-1}^{2\vert l\vert -2}\right)}{2^{2\vert l\vert -2}} \ \ \ \ (g\rightarrow +\infty).
\end{equation}
The large $l$ behavior of\ (\ref{eq4.1.4}) can be determined by the singular term $1/\sqrt{1-s}$ on the right-hand side of\ (\ref{eq4.1.3}) through the appropriate Tauberian theorem (here Darboux's theorem\ \cite{Hen}, see also e.g. Theorem VI.1 in \cite{FS}). A factor $1/\sqrt{1-s}$,\ ($s\rightarrow 1$), in $\tilde{p}(g,s)$ translates into a factor $1/\sqrt{\pi l}$,\ ($l\rightarrow +\infty$), in $p(g,l)$ and one gets
\begin{equation}\label{eq4.1.5}
p(g,l)\sim\frac{f(g)I_2(g)}{\sqrt{\pi l}} \ \ \ \ (g\rightarrow +\infty\ {\rm then}\ l\rightarrow +\infty).
\end{equation}
To go beyond Eqs.\ (\ref{eq4.1.4}) and\ (\ref{eq4.1.5}) we need to specify $f(\eta)$ further on, which we will now do by studying a particular case of class A jumps.
%
%
\subsubsection{Jump distribution with an algebraic tail: scaling form}\label{sec4.1.1}
\begin{figure}[htbp]
\begin{center}
\includegraphics [width=9cm] {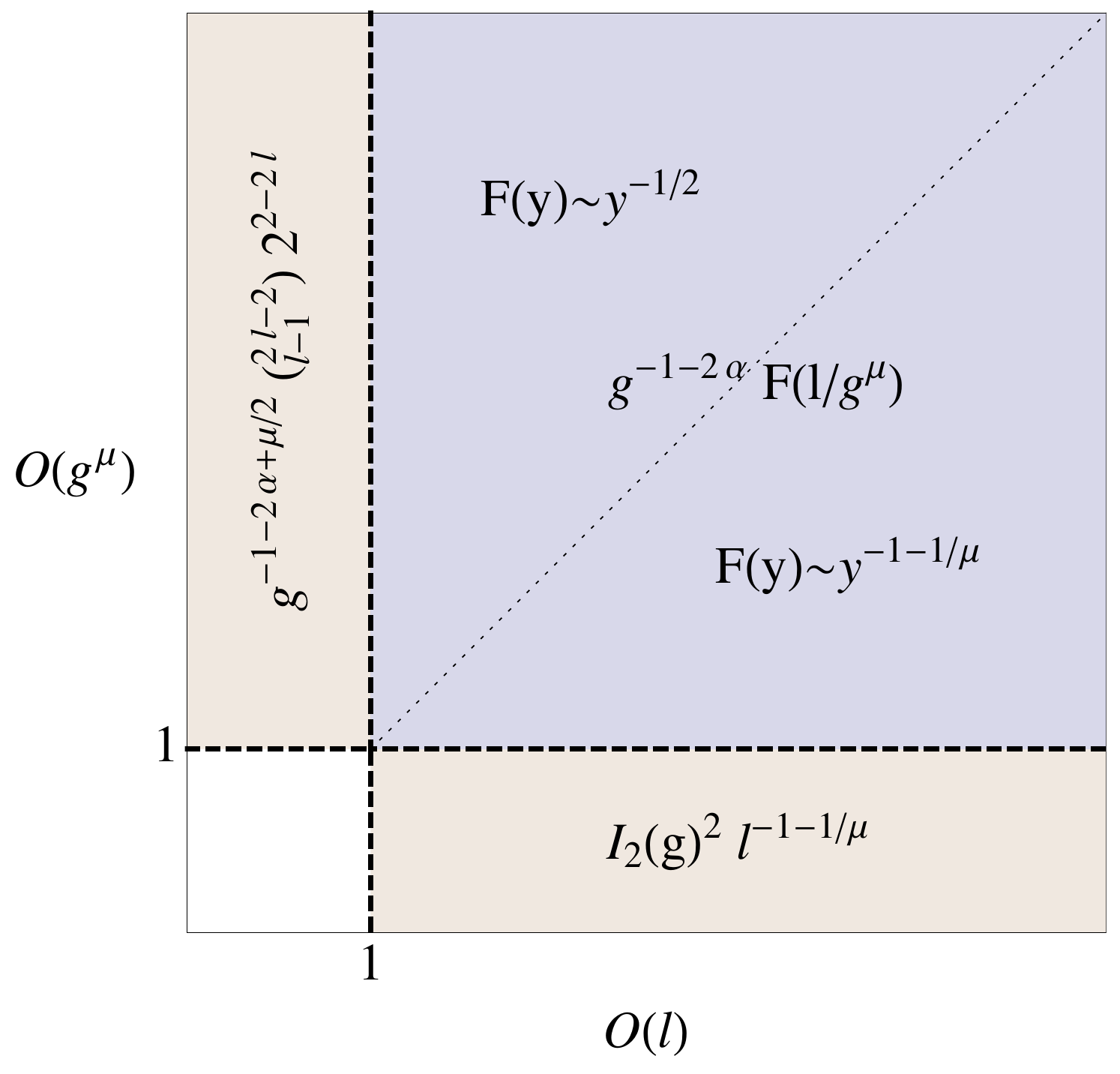}
\caption{\textsl{\it Schematic representation of the asymptotic behaviors of $p(g,l)$ for jump distributions with an algebraic tail of the form\ (\ref{eq4.1.1.1}) (amplitudes are given in the text). For $l\gg 1$ and $g\lesssim O(1)$, $p(g,l)$ is given by Eq.\ (\ref{eq4.13}). For both $l\gg 1$ and $g\gg 1$, one has the scaling form\ (\ref{eq4.1.1.8}) with $F(y)$ respectively given by\ (\ref{eq4.1.1.10}) if $l\gg g^\mu$ and by\ (\ref{eq4.1.1.11}) if $l\ll g^\mu$. For $l=O(1)$ and $g\gg 1$, $p(g,l)$ is given by Eq.\ (\ref{eq4.1.1.3}).}}
\label{fig2}
\end{center}
\end{figure}

Here we consider class A jumps the distribution of which has an algebraic tail of the form
\begin{equation}\label{eq4.1.1.1}
f(\eta)\sim \frac{C_\alpha d^\alpha}{\eta^{\alpha +1}} \ \ \ \ (\eta\rightarrow +\infty),
\end{equation}
with $0<\alpha\ne 2$ and $\mu =\min(2,\alpha)$. Here $C_\alpha$ is a constant and $d$ a length scale not necessarily equal to $a$. In this case, the large $g$ behavior of $I_2(g)$ can be computed explicitly along the same line as the one leading to\ (\ref{eq3.20}). After some straightforward algebra one obtains
\begin{eqnarray}\label{eq4.1.1.2}
I_2(g)&\sim&\frac{C_\alpha}{\Gamma(\mu/2+1)}\left(\frac{d^\alpha}{a^{\mu/2}}\right)
\frac{1}{g^{\alpha -\mu/2}}\int_0^{+\infty}\frac{\overline{x}^{\mu/2}}{(1+\overline{x})^{\alpha +1}}\, 
d\overline{x} \ \ \ \ (g\rightarrow +\infty) \nonumber \\
&=&
\frac{C_\alpha \Gamma(\alpha -\mu/2)}{\Gamma(\alpha +1)}\left(\frac{d^\alpha}{a^{\mu/2}}\right)
\frac{1}{g^{\alpha -\mu/2}},
\end{eqnarray}
where the integral over $\overline{x}$ in the first line is equal to $\Gamma(\mu/2+1)\Gamma(\alpha -\mu/2)/\Gamma(\alpha +1)$. Equations\ (\ref{eq4.1.4}) and\ (\ref{eq4.1.5}) in which $f(g)$ is given by\ (\ref{eq4.1.1.1}) and $I_2(g)$ by\ (\ref{eq4.1.1.2}) read, [we recall that $\mu$ is related to $\alpha$ by $\mu =\min(2,\alpha)$],
\begin{equation}\label{eq4.1.1.3}
p(g,l)\sim\left(\frac{d^{2\alpha}}{a^{\mu/2}}\right)
\frac{\sqrt{\pi}B_{\alpha}}{g^{1+2\alpha -\mu/2}}
\, \frac{\left(_{\vert l\vert -1}^{2\vert l\vert -2}\right)}{2^{2\vert l\vert -2}} \ \ \ \ (g\rightarrow +\infty),
\end{equation}
and
\begin{equation}\label{eq4.1.1.4}
p(g,l)\sim\left(\frac{d^{2\alpha}}{a^{\mu/2}}\right)
\frac{B_{\alpha}}{l^{1/2}g^{1+2\alpha -\mu/2}} \ \ \ \ (g\rightarrow +\infty\ {\rm then}\ l\rightarrow +\infty),
\end{equation}
respectively, with
\begin{equation}\label{eq4.1.1.5}
B_{\alpha}=\frac{C_\alpha^2 \Gamma(\alpha -\mu/2)}{\sqrt{\pi}\Gamma(\alpha +1)}.
\end{equation}
It may be interesting to compare\ (\ref{eq4.1.1.4}) with the $g\rightarrow +\infty$ limit of\ (\ref{eq4.13}) in which $I_2(g)$ is given by\ (\ref{eq4.1.1.2}):
\begin{equation}\label{eq4.1.1.6}
p(g,l)\sim\left(\frac{d^{2\alpha}}{a^{\mu +1}}\right)
\frac{A_{\alpha}}{l^{1+1/\mu}g^{2\alpha -\mu}} \ \ \ \ (l\rightarrow +\infty\ {\rm then}\ g\rightarrow +\infty),
\end{equation}
with
\begin{equation}\label{eq4.1.1.7}
A_{\alpha}=\frac{C_\alpha^2 \Gamma(1+1/\mu)\Gamma(\alpha -\mu/2)^2}
{\pi\Gamma(\alpha +1)^2}.
\end{equation}
Expressions\ (\ref{eq4.1.1.4}) and\ (\ref{eq4.1.1.6}) suggest the existence of a scaling form $p(g,l)\sim g^{-1-2\alpha}F(l/g^\mu)$ when both $l$ and $g$ are large, whatever their relative size, with $F(y)\sim 1/\sqrt{y}$, ($y\rightarrow 0$), and $F(y)\sim 1/y^{1+1/\mu}$, ($y\rightarrow +\infty$). It is shown in Appendix\ \ref{app4} that such a scaling form does exist in the whole domain of large $l$ and $g$. Namely, one finds
\begin{equation}\label{eq4.1.1.8}
p(g,l)\sim\frac{d^{2\alpha}}{g^{1+2\alpha}}F_{\alpha}\left(\frac{a^\mu l}{g^\mu}\right)
 \ \ \ \ (l,g\rightarrow +\infty),
\end{equation}
where the scaling function $F_{\alpha}(y)$ can be written in the integral form
\begin{eqnarray}\label{eq4.1.1.9}
F_{\alpha}(y)&=&\frac{\sqrt{\pi} B_{\alpha}}{y}
\int_{\cal{L}}\frac{d\overline{p}}{2i\pi}{\rm e}^{\overline{p}}
\int_0^{+\infty}\frac{d\overline{x}}{(1+\overline{x})^{\alpha +1}} \\
&\times&\int_{\cal{L}}
\frac{d\overline{\lambda}}{2i\pi}\, \frac{\rm{e}^{\overline{\lambda}\overline{x}}}
{\overline{\lambda}^{\mu/2}}\, 
\exp\left\lbrack-\frac{\overline{\lambda}}{\pi}
\int_0^{+\infty}\frac{1}{q^2+\overline{\lambda}^2}\, 
\ln\left(1+\frac{\overline{p}}{yq^\mu}\right)\, dq\right\rbrack , \nonumber
\end{eqnarray}
with large and small argument behaviors respectively given by
\begin{equation}\label{eq4.1.1.10}
F_{\alpha}(y)\sim\frac{A_{\alpha}}{y^{1+1/\mu}}
 \ \ \ \ (y\rightarrow +\infty),
\end{equation}
and
\begin{equation}\label{eq4.1.1.11}
F_{\alpha}(y)\sim\frac{B_{\alpha}}{\sqrt{y}}
 \ \ \ \ (y\rightarrow 0).
\end{equation}
The interested reader is referred to Appendix\ \ref{app4} for details. Figure\ \ref{fig2} summarizes the different behaviors of $p(g,l)$ in the plane $(l,g^\mu)$ for jump distributions with an algebraic tail of the form\ (\ref{eq4.1.1.1}).
\bigskip

As a first application of the scaling form\ (\ref{eq4.1.1.8}) we can recover the large $g$ behavior\ (\ref{eq3.23}) of $p_{\rm gap}(g)$ in a very simple way, as we will now see. Fix $\Lambda\gg 1$ and write
\begin{equation}\label{eq4.1.1.12}
p_{\rm gap}(g)=2\sum_{l=1}^{+\infty}p(g,l)
=2\sum_{l=1}^{\Lambda -1}p(g,l) +2\sum_{l=\Lambda}^{+\infty}p(g,l).
\end{equation}
For large $g$, it follows from\ (\ref{eq4.1.1.10}) and\ (\ref{eq4.1.1.11}) that the sum over $l$ in\ (\ref{eq4.1.1.12}) is determined by large values of $l\sim g^\mu$ and one has
\begin{eqnarray}\label{eq4.1.1.13}
&&p_{\rm gap}(g)\sim 2\sum_{l=\Lambda}^{+\infty}
\frac{d^{2\alpha}}{g^{1+2\alpha}}F_{\alpha}\left(\frac{a^\mu l}{g^\mu}\right)
=\frac{2d^{2\alpha}}{g^{1+2\alpha-\mu}}\sum_{l=\Lambda}^{+\infty}
F_{\alpha}\left(\frac{a^\mu l}{g^\mu}\right)\frac{1}{g^\mu} \nonumber \\
&&\sim\frac{2d^{2\alpha}}{a^\mu g^{1+2\alpha-\mu}}
\int_{\Lambda a^\mu/g^\mu}^{+\infty}F_{\alpha}(y)\, dy
\sim\frac{2d^{2\alpha}}{a^\mu g^{1+2\alpha-\mu}}
\int_0^{+\infty}F_{\alpha}(y)\, dy
 \ \ \ \ (g\rightarrow +\infty),
 \end{eqnarray}
which is nothing but Eq.\ (\ref{eq3.23}) with $\tilde{C}_\alpha$ given in terms of $F_\alpha$ as
\begin{equation}\label{eq4.1.1.14}
\tilde{C}_{\alpha}=2\int_0^{+\infty}F_{\alpha}(y)\, dy.
\end{equation}
%
%
\subsubsection{L\'evy flights ($0<\mu <2$)}\label{sec4.1.2}
For $0<\mu <2$ the variance of the jump distribution does not exist and the random walk is a L\'evy flight. As mentioned in Sec.\ \ref{sec3.2}, the large $\eta$ behavior of $f(\eta)$ in this case can be obtained from the small $k$ behavior of its Fourier transform $\hat{f}(k)$ (see Appendix\ \ref{app2}). One gets the Eq.\ (\ref{eq3.18}),
\begin{equation*}
f(\eta)\sim \sin\left(\frac{\mu\pi}{2}\right)\Gamma(\mu+1)
\frac{a^\mu}{\pi \eta^{\mu+1}} \ \ \ \ (\eta\rightarrow +\infty),
\end{equation*}
which is an algebraic tail of the form\ (\ref{eq4.1.1.1}) with $0<\alpha = \mu <2$, $d=a$, and $C_\mu=\pi^{-1}\sin(\mu\pi/2)\Gamma(\mu +1)$. The asymptotic behavior of $p(g,l)$ for large $g$ or $l$ (or both) is thus merely obtained as a particular case of the preceding analysis. For the reader's convenience we summarize here the results, giving all the parameters and amplitudes as functions of $\mu$ and $a$.

Beside Eq.\ (\ref{eq4.13}) which holds as it is, Eq.\ (\ref{eq4.1.1.3}) reads
\begin{equation}\label{eq4.1.2.1}
p(g,l)\sim a^{3\mu/2}
\frac{\sqrt{\pi}B_\mu}{g^{1+3\mu/2}}
\, \frac{\left(_{\vert l\vert -1}^{2\vert l\vert -2}\right)}{2^{2\vert l\vert -2}} \ \ \ \ (g\rightarrow +\infty),
\end{equation}
with
\begin{equation}\label{eq4.1.2.2}
B_\mu =
\frac{\Gamma(1+\mu)}{\sqrt{\pi}\, \Gamma(\mu/2)\Gamma(1-\mu/2)^2},
\end{equation}
where we have used the reflection formula $\Gamma(z)\Gamma(1-z)=\pi/\sin(\pi z)$. The scaling form\ (\ref{eq4.1.1.8}) reads
\begin{equation}\label{eq4.1.2.3}
p(g,l)\sim\frac{a^{2\mu}}{g^{1+2\mu}}F_\mu\left(\frac{a^\mu l}{g^\mu}\right)
 \ \ \ \ (l,g\rightarrow +\infty),
\end{equation}
with
\begin{equation}\label{eq4.1.2.4}
F_\mu(y)\sim\frac{A_\mu}{y^{1+1/\mu}}
 \ \ \ \ (y\rightarrow +\infty),
\end{equation}
where
\begin{equation}\label{eq4.1.2.5}
A_\mu =
\frac{\Gamma(1+1/\mu)}{\pi\, \Gamma(1-\mu/2)^2},
\end{equation}
and
\begin{equation}\label{eq4.1.2.6}
F_\mu(y)\sim\frac{B_\mu}{\sqrt{y}}
 \ \ \ \ (y\rightarrow 0).
\end{equation}
As already mentioned at the end of Sec.\ \ref{sec3}, the large $g$ behavior\ (\ref{eq3.23}) of $p_{\rm gap}(g)$ with $0<\alpha = \mu <2$ and $d=a$ coincides with Eq.\ (\ref{eq3.21}), as it should be.
%
%
\subsection{Class B jumps: exponentially decreasing $\bm{f(\eta)}$}\label{sec4.2}
\begin{figure}[htbp]
\begin{center}
\includegraphics [width=9cm] {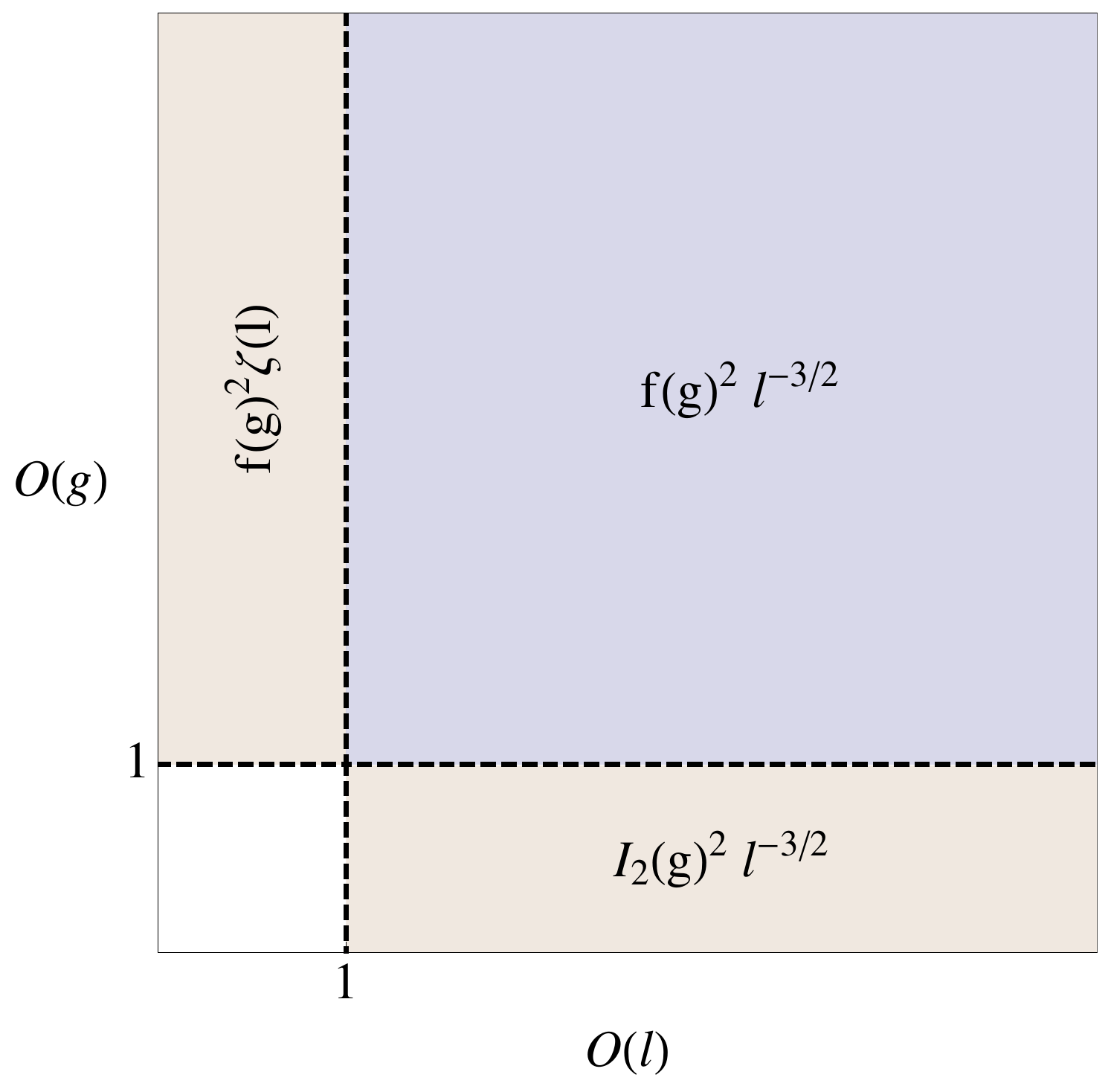}
\caption{\textsl{\it Schematic representation of the asymptotic behaviors of $p(g,l)$ for class B jumps (amplitudes are given in the text). For $l\gg 1$ and $g\lesssim O(1)$, $p(g,l)$ is given by Eq.\ (\ref{eq4.13}). For both $l\gg 1$ and $g\gg 1$, it reduces to Eq.\ (\ref{eq4.2.7}) whatever the relative sizes of $l$ and $g$. For $l=O(1)$ and $g\gg 1$, $p(g,l)$ is given by Eq.\ (\ref{eq4.2.5}).}}
\label{fig3}
\end{center}
\end{figure}

Class B jumps are defined by
\begin{equation}\label{eq4.2.1}
f(g+x)\sim f(g)\exp(-cx) \ \ \ \ (g\rightarrow +\infty),
\end{equation}
for some $c>0$ and any fixed $x$. The two jump distributions considered in Sec.\ \ref{sec3.1} are class B. Substituting\ (\ref{eq4.2.1}) for $f(g+x)$ on the right-hand side of\ (\ref{eq2.17}) and using\ (\ref{eq2.11}), one finds
\begin{equation}\label{eq4.2.2}
I_1(g,s)\sim f(g)\, s\phi(c,s) \ \ \ \ (g\rightarrow +\infty),
\end{equation}
and
\begin{equation}\label{eq4.2.3}
I_2(g)\sim f(g)\, \frac{\phi(c,1)}{c} \ \ \ \ (g\rightarrow +\infty),
\end{equation}
which, together with Eq.\ (\ref{eq2.16}), yields
\begin{equation}\label{eq4.2.4}
\tilde{p}(g,s)\sim \frac{\phi(c,1)}{c}\, f(g)^2\, s\phi(c,s) \ \ \ \ (g\rightarrow +\infty).
\end{equation}
For the two jump distributions considered in Sec.\ \ref{sec3.1}, it can be readily seen that\ (\ref{eq4.2.4}) with $s=1$ gives the same large $g$ behavior of $p_{\rm gap}(g)=2\tilde{p}(g,1)$ as what we get from the large $g$ limit of\ (\ref{eq3.6}) or\ (\ref{eq3.12}), as it should be. Now, writing $\zeta(l)=\lbrack s^l\rbrack\, s\phi(c,s)$, the term of order $l$ in the expansion of $s\phi(c,s)$ as a power series of $s$, it results from\ (\ref{eq4.2.4}) that
\begin{equation}\label{eq4.2.5}
p(g,l)\sim \frac{\phi(c,1)}{c}\, f(g)^2\, \zeta(\vert l\vert ) \ \ \ \ (g\rightarrow +\infty).
\end{equation}
The large $l$ behavior of\ (\ref{eq4.2.5}) can be deduced from the behavior of\ (\ref{eq4.2.4}) near its dominant singularity at $s=1$. Eq.\ (\ref{eq4.8}) with $\mu =2$ and $\lambda =c$ leads to
\begin{equation}\label{eq4.2.6}
\tilde{p}(g,s)\sim -\frac{1}{a}\left\lbrack\frac{\phi(c,1)}{c}\right\rbrack^2
f(g)^2 s\sqrt{1-s} \ \ \ \ (g\rightarrow +\infty\ {\rm then}\ s\rightarrow 1),
\end{equation}
and by Darboux's theorem\ \cite{Hen} (see also e.g. Theorem VI.1 in \cite{FS}), one obtains
\begin{equation}\label{eq4.2.7}
p(g,l)\sim \frac{1}{2\sqrt{\pi}\, a}\left\lbrack\frac{\phi(c,1)}{c}\right\rbrack^2
\frac{f(g)^2}{l^{3/2}} \ \ \ \ (g\rightarrow +\infty\ {\rm then}\ l\rightarrow +\infty).
\end{equation}
Interestingly enough, Eq.\ (\ref{eq4.13}) with $I_2(g)$ given by\ (\ref{eq4.2.3}) and $\mu =2$ yields the same expression\ (\ref{eq4.2.7}) but with the $g$ and $l$ limits interchanged. It means that, for class B jumps, the behavior of $p(g,l)$ for both $g$ and $l$ large is correctly given by\ (\ref{eq4.2.7}) whatever the order in which the two limits are taken. Figure\ \ref{fig3} summarizes the different behaviors of $p(g,l)$ in the plane $(l,g)$ for class B jumps.
\bigskip
%
%
%
\subsection{Class C jumps: fast decreasing $\bm{f(\eta)}$ and concentration of $\bm{p(g,l)}$ onto $\bm{l=\pm 1}$}\label{sec4.3}
Class C jumps are defined by
\begin{equation}\label{eq4.3.1}
f(g+x)\sim f(g)\exp(-cg^\delta x+\theta(x,g)) \ \ \ \ (g\rightarrow +\infty),
\end{equation}
for some $c,\delta>0$, any fixed $x$, and where $\theta(x,g)$ is such that $\lim_{g\rightarrow +\infty}\theta(x\lesssim g^{-\delta},g)=0$. Super-exponentially distributed jumps are class C. Substituting\ (\ref{eq4.3.1}) for $f(g+x)$ on the right-hand side of\ (\ref{eq2.17}) and using\ (\ref{eq2.11}), one finds
\begin{eqnarray}\label{eq4.3.2}
I_1(g,s)&\sim& f(g)\, s\int_0^{+\infty}u(x,s){\rm e}^{\theta(x,g)}
{\rm e}^{-cg^\delta x}dx \nonumber \\
&\sim& f(g)\, s\int_0^{+\infty}u(x,s){\rm e}^{-cg^\delta x}dx \nonumber \\
&=&f(g)\, s\phi(cg^\delta ,s) \ \ \ \ (g\rightarrow +\infty),
\end{eqnarray}
where we have used $\exp\theta(x,g)\sim 1$ as $g\rightarrow +\infty$ for $x$ in the effective domain of integration $x\lesssim g^{-\delta}$. Then, from the large $\lambda$ limit of the expression\ (\ref{eq2.8}) for $\phi(\lambda ,s)$ one obtains
\begin{equation}\label{eq4.3.3}
I_1(g,s)\sim f(g)\, \left(
s-\frac{s}{\pi cg^\delta}\int_0^{+\infty}\ln\lbrack 1-s\hat{f}(k)\rbrack\, dk\right)
 \ \ \ \ (g\rightarrow +\infty).
\end{equation}
Similarly, one has
\begin{eqnarray}\label{eq4.3.4}
I_2(g)&\sim& f(g)\int_0^{+\infty}h(x,1){\rm e}^{\theta(x,g)}
{\rm e}^{-cg^\delta x}dx \nonumber \\
&\sim& f(g)\int_0^{+\infty}h(x,1){\rm e}^{-cg^\delta x}dx \nonumber \\
&=&f(g)\frac{\phi(cg^\delta ,1)}{cg^\delta}
\sim\frac{f(g)}{cg^\delta} \ \ \ \ (g\rightarrow +\infty).
\end{eqnarray}
Eqs.\ (\ref{eq2.16}),\ (\ref{eq4.3.3}), and\ (\ref{eq4.3.4}) yield
\begin{equation}\label{eq4.3.5}
\tilde{p}(g,s)\sim\frac{f(g)^2}{cg^\delta}\, \left(
s-\frac{s}{\pi cg^\delta}\int_0^{+\infty}\ln\lbrack 1-s\hat{f}(k)\rbrack\, dk\right)
 \ \ \ \ (g\rightarrow +\infty),
\end{equation}
from which one readily obtains, by expanding the logarithm in power series of $s$,
\begin{equation}\label{eq4.3.6}
p(g,l)\sim\left\lbrace
\begin{array}{lr}
c^{-1}f(g)^2 g^{-\delta}&l=\pm 1,\\
c^{-2}f(g)^2 g^{-2\delta}p(0,\vert l\vert -1\vert 0,0)/(\vert l\vert -1)&\vert l\vert\ge 2,
\end{array}\right.
 \ \ \ \ (g\rightarrow +\infty),
\end{equation}
where we have rewritten the integrals over $k$ as
\begin{equation}\label{eq4.3.7}
\frac{1}{\pi}\int_0^{+\infty}\hat{f}(k)^n dk=
\frac{1}{2\pi}\int_{-\infty}^{+\infty}\hat{f}(k)^n dk=
p(0,n\vert 0,0).
\end{equation}
An immediate, interesting, consequence of\ (\ref{eq4.3.6}) is the concentration of $p(g,l)$ onto $l=\pm 1$ as $g$ gets large in the sense that, for any $l$ with $\vert l\vert\ge 2$, Eq.\ (\ref{eq4.3.6}) entails
\begin{equation}\label{eq4.3.8}
\frac{p(g,l)}{p(g,\pm 1)}\sim \frac{p(0,\vert l\vert -1\vert 0,0)}{c(\vert l\vert -1)}
\, \frac{1}{g^{\delta}}\rightarrow 0,
\end{equation}
as $g\rightarrow +\infty$. Practically, this result means that for class C jumps a large gap is mainly due to configurations with adjacent first and second maxima. Other configurations necessarily involve (at least) one jump larger than the gap the contribution of which gets rapidly negligible as the gap gets large due to the very fast decrease of $f(\eta)$ for large $\eta$. We can take advantage of this concentration to get the large $g$ behavior of $p_{\rm gap}(g)$ in a very simple way. Indeed, if $f(\eta)$ is fast decreasing it follows immediately from\ (\ref{eq4.3.8}) and the first line of\ (\ref{eq4.3.6}) that $p_{\rm gap}(g)\sim p(g,\vert l\vert =1)\sim 2c^{-1}f(g)^2 g^{-\delta}$ as $g\rightarrow +\infty$. Note that we expect a similar concentration onto $l=\pm 1$ if $f(\eta)$ has a bounded support $-\eta_{max}\le\eta\le\eta_{max}$ and $g\rightarrow +\infty$ is replaced with $g\rightarrow\eta_{max}$, the largest possible value of the gap. A detailed study of this interesting case (where neither $g$ nor $l$ are large) is left for a future work.

\begin{figure}[htbp]
\begin{center}
\includegraphics [width=9cm] {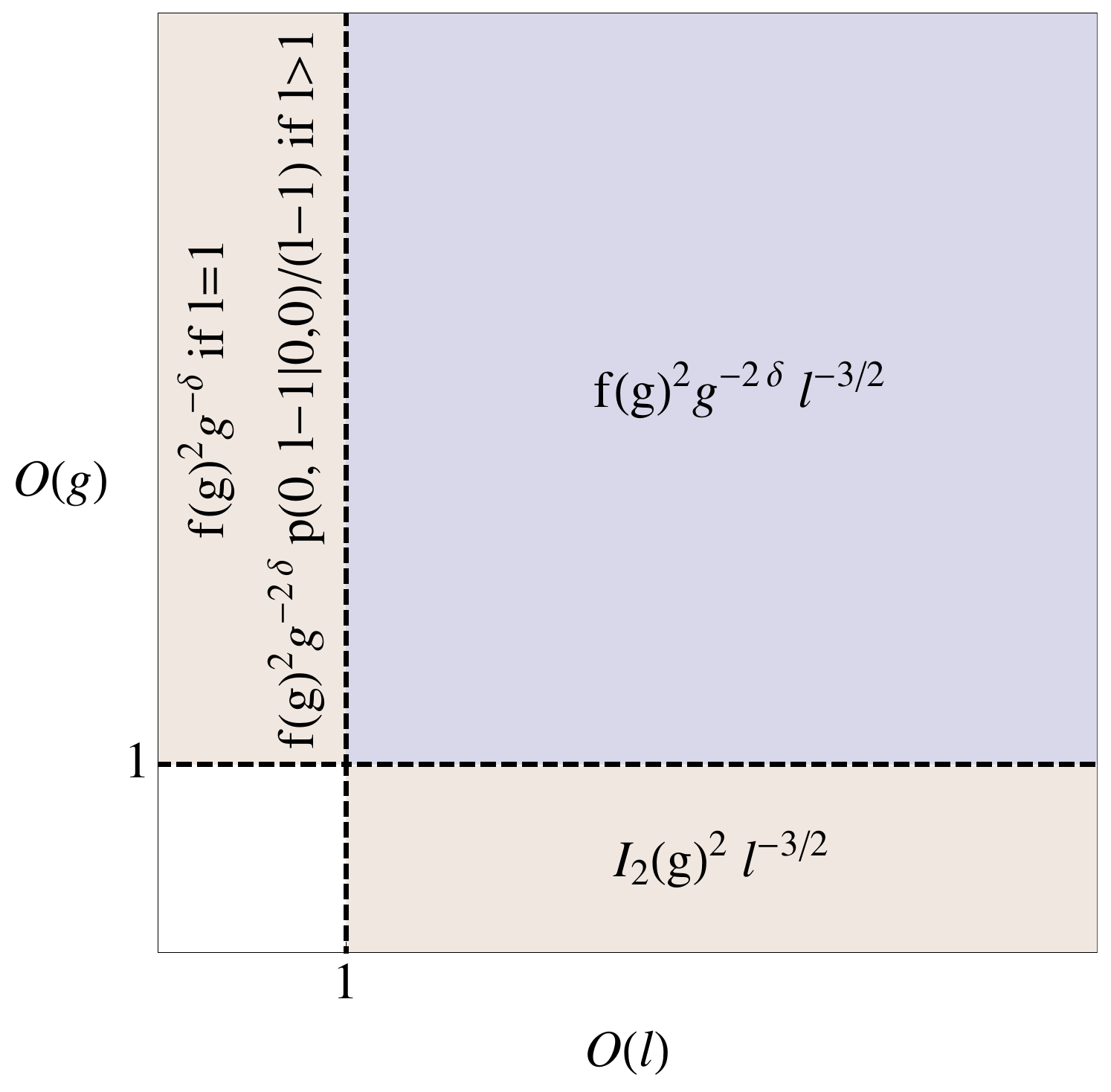}
\caption{\textsl{\it Schematic representation of the asymptotic behaviors of $p(g,l)$ for class C jumps (amplitudes are given in the text). For $l\gg 1$ and $g\lesssim O(1)$, $p(g,l)$ is given by Eq.\ (\ref{eq4.13}). For both $l\gg 1$ and $g\gg 1$, it reduces to Eq.\ (\ref{eq4.3.10}) whatever the relative sizes of $l$ and $g$. For $l=O(1)$ and $g\gg 1$, $p(g,l)$ is given by Eq.\ (\ref{eq4.3.6}).}}
\label{fig4}
\end{center}
\end{figure}

The large $l$ behavior of\ (\ref{eq4.3.6}) follows straightforwardly from the large $n$ behavior of\ (\ref{eq4.3.7}) which reads, for $\mu =2$ [see Eq.\ (\ref{eq6.10})],
\begin{equation}\label{eq4.3.9}
p(0,n\vert 0,0)\sim\frac{1}{2\sqrt{\pi}\, a}\, \frac{1}{\sqrt{n}} \ \ \ \ (n\rightarrow +\infty),
\end{equation}
and one gets
\begin{equation}\label{eq4.3.10}
p(g,l)\sim\frac{1}{2\sqrt{\pi}\, ac^2}\, \frac{f(g)^2}{l^{3/2}g^{2\delta}}
 \ \ \ \ (g\rightarrow +\infty\ {\rm then}\ l\rightarrow +\infty).
 \end{equation}
It can be checked that Eq.\ (\ref{eq4.13}) with $I_2(g)$ given by\ (\ref{eq4.3.4}) and $\mu =2$ gives the same expression\ (\ref{eq4.3.10}) with $g$ and $l$ limits interchanged. Thus, just like what we found for class B jumps, for class C jumps too the behavior of $p(g,l)$ for both $g$ and $l$ large is correctly given by\ (\ref{eq4.3.10}) whatever the order in which the two limits are taken. Figure\ \ref{fig4} summarizes the different behaviors of $p(g,l)$ in the plane $(l,g)$ for class C jumps.
\bigskip
%
%
%
\section{Large $\bm{l}$ behavior of the marginal distribution $\bm{p_{\rm time}(l)}$}\label{sec5}
We now have all the necessary ingredients to determine the large $l$ behavior of the marginal distribution $p_{\rm time}(l)$ of the time between the first two maxima. To this end, fix $\Lambda_1\gg 1$, $\Lambda_2 =O(1)$, and for $l>(\Lambda_1/\Lambda_2)^\mu$ write
\begin{equation}\label{eq5.1}
p_{\rm time}(l)=\int_0^{+\infty}p(g,l)\, dg
=\int_0^{\Lambda_1}p(g,l)\, dg+\int_{\Lambda_1}^{\Lambda_2 l^{1/\mu}}p(g,l)\, dg
+\int_{\Lambda_2 l^{1/\mu}}^{+\infty}p(g,l)\, dg.
\end{equation}
For $1<\mu\le 2$ and $l\rightarrow +\infty$,\ (\ref{eq5.1}) is dominated by the first two integrals in which one can use the large $l$ expression\ (\ref{eq4.13}) of $p(g,l)$. One finds
\begin{eqnarray}\label{eq5.2}
p_{\rm time}(l)&\sim&
\frac{\Gamma(1+1/\mu)}{\pi al^{1+1/\mu}}\int_0^{\Lambda_2 l^{1/\mu}} I_2(g)^2\, dg \nonumber \\
&\sim&
\frac{\Gamma(1+1/\mu)}{\pi al^{1+1/\mu}}\int_0^{+\infty} I_2(g)^2\, dg
 \ \ \ \ (l\rightarrow +\infty).
\end{eqnarray}
For $0<\mu <1$ and $l\rightarrow +\infty$,\ (\ref{eq5.1}) is dominated by the contribution of large, $O(l^{1/\mu})$, values of $g$, i.e. by the last two integrals in which one can use the large $l$ and $g$ scaling form\ (\ref{eq4.1.2.3}) of $p(g,l)$. One gets
\begin{eqnarray}\label{eq5.3}
p_{\rm time}(l)&\sim&\int_{\Lambda_1}^{+\infty}
\frac{a^{2\mu}}{g^{1+2\mu}}F_\mu\left(\frac{a^\mu l}{g^\mu}\right)\, dg \nonumber \\
&=&\frac{1}{\mu l^2}\int_0^{a^\mu l/\Lambda_1^\mu} yF_\mu(y)\, dy \nonumber \\
&\sim&\frac{1}{\mu l^2}\int_0^{+\infty} yF_\mu(y)\, dy
 \ \ \ \ (l\rightarrow +\infty),
\end{eqnarray}
where we have made the change of variable $y=a^\mu l/g^\mu$. Finally, for  $\mu =1$ and $l\rightarrow +\infty$, the $g$-integral\ (\ref{eq5.1}) is again dominated by the contribution of large values of $g$ and from the large argument behavior\ (\ref{eq4.1.2.4}) of $F_\mu (y)$ (with $\mu =1$) it follows
\begin{eqnarray}\label{eq5.4}
p_{\rm time}(l)&\sim&\int_{\Lambda_1}^{+\infty}
\frac{a^{2}}{g^3}F_1\left(\frac{al}{g}\right)\, dg
=\frac{1}{l^2}\int_0^{al/\Lambda_1} yF_\mu(y)\, dy \nonumber \\
&\sim&\frac{1}{\pi^2 l^2}\int_0^{al/\Lambda_1} \frac{dy}{y}
\sim\frac{1}{\pi^2}\, \frac{\ln(l)}{l^2} \ \ \ \ (l\rightarrow +\infty),
\end{eqnarray}
(where $y=al/g$). This asymptotic behavior could equivalently have been obtained from the large $l$ expression\ (\ref{eq4.13}) of $p(g,l)$ as
\begin{equation*}
p_{\rm time}(l)\sim\frac{1}{\pi al^2}\int_0^{\Lambda_2 l}I_2(g)^2\, dg
\sim\frac{1}{\pi^2 l^2}\int_0^{\Lambda_2 l}\frac{dg}{g}
\sim\frac{1}{\pi^2}\, \frac{\ln(l)}{l^2} \ \ \ \ (l\rightarrow +\infty).
\end{equation*}
where we have used $I_2(g)\sim\sqrt{a/\pi g}$, ($g\rightarrow +\infty$), which follows from\ (\ref{eq4.1.1.2}) with $\alpha =\mu =1$, $d=a$, and $C_1 =1/\pi$.

To summarize, we find three different regimes for the large $l$ behavior of the marginal distribution $p_{\rm time}(l)$, depending on the value of $\mu$:
\begin{equation}\label{eq5.5}
p_{\rm time}(l)\sim\left\lbrace
\begin{array}{ll}
\mathcal{A}_I l^{-1-1/\mu}&1<\mu\le 2 \\
\mathcal{A}_{II} l^{-2}\ln(l)&\mu =1 \\
\mathcal{A}_{III}l^{-2}&0<\mu <1
\end{array}\right.
 \ \ \ \ (l\rightarrow +\infty),
\end{equation}
where the amplitudes $\mathcal{A}_I$, $\mathcal{A}_{II}$, and $\mathcal{A}_{III}$ are respectively given by
\begin{equation}\label{eq5.6}
\begin{array}{l}
\mathcal{A}_I =(\pi a)^{-1}\Gamma(1+1/\mu)\int_0^{+\infty}I_2(g)^2\, dg, \\
\mathcal{A}_{II} =\pi^{-2}, \\
\mathcal{A}_{III} =\mu^{-1}\int_0^{+\infty} yF_\mu(y)\, dy.
\end{array}
\end{equation}
Note that, unlike $\mathcal{A}_I$, both $\mathcal{A}_{II}$ and $\mathcal{A}_{III}$ are independent of $a$. The third line of Eq.\ (\ref{eq5.5}) reveals an unexpected freezing phenomenon of the exponent characterizing the algebraic tail of $p_{\rm time}(l)$ as $\mu$ decreases past the critical value $\mu_c =1$. In particular, it follows immediately from\ (\ref{eq5.5}) that the first moment of $p_{\rm time}(l)$ is never defined. This means that although the typical size of $L_n$ is $O(1)$, where $n$ is the walk total duration, its average always diverges with $n$, and\ (\ref{eq5.5}) yields the estimates
\begin{equation}\label{eq5.7}
\langle\vert L_n\vert\rangle\sim\left\lbrace
\begin{array}{ll}
n^{1-1/\mu}&1<\mu\le 2 \\
\ln(n)^2&\mu =1 \\
\ln(n)&0<\mu <1
\end{array}\right.
 \ \ \ \ (n\rightarrow +\infty).
\end{equation}
%
%
\section{Generating function of $\bm{p(g,l)}$ for a bridge}\label{sec6}
In the conclusion of Sec.\ \ref{sec2} we mentioned the possibility that $p(g,l)$ might not be affected by a boundary condition at the the end of the walk. In this section we show that this is actually so in the case where the free-end walk considered so far is replaced with a bridge in which the walker is conditioned to return to $x=0$ at the end of the walk. With this constraint, and assuming without loss of generality that $l>0$, one finds that Eq.\ (\ref{eq2.1}) modifies to
\begin{eqnarray}\label{eq6.1}
p_n^{br}(g,l_1,l,l_3)&=&\left\lbrack\int_0^{+\infty}p_{<z}(z,l_1\vert 0,0)
\left(\int_{y<z}p_{<z}(y,l-1\vert z,0)f(z+g-y)\, dy\right)\right. \nonumber \\
&\times&\left.\left(\int_{x<z}f(z+g-x)p_{<z}(0,l_3-1\vert x,0)\, dx\right)\, dz
\right\rbrack\, \frac{\delta_{l_1+l+l_3,n}}{p(0,n\vert 0,0)},
\end{eqnarray}
where the superscript $br$ stands for ``bridge". According to Eq.\ (\ref{eq2.3}), the first parenthesis is equal to $w_2(g,l)$, independent of $z$. In the second parenthesis we write $p_{<z}(0,l_3-1\vert x,0)=p_{>0}(z,l_3-1\vert z-x,0)$ (see Fig. \ref{fig_propag} a)), obtained by taking $z$ as a new origin and reversing space direction, which gives
\begin{eqnarray}\label{eq6.2}
&&\int_{x<z}f(z+g-x)p_{<z}(0,l_3-1\vert x,0)\, dx= \nonumber \\
&&\int_{x<z}f(z+g-x)p_{>0}(z,l_3-1\vert z-x,0)\, dx= \nonumber \\
&&\int_{x>0}f(g+x)p_{>0}(z,l_3-1\vert x,0)\, dx,
\end{eqnarray}
where we have made the change of variable $x\rightarrow z-x$ in the third line. Writing then $p_{<z}(z,l_1\vert 0,0)=p_{>0}(0,l_1\vert z,0)$, obtained by taking $z$ as a new origin and reversing space direction, one has
\begin{eqnarray}\label{eq6.3}
&&\int_0^{+\infty}p_{<z}(z,l_1\vert 0,0)
\left(\int_{x<z}f(z+g-x)p_{<z}(0,l_3-1\vert x,0)\, dx\right)\, dz= \nonumber \\
&&\int_0^{+\infty}f(g+x)\left(\int_0^{+\infty}p_{<z}(z,l_1\vert 0,0)
p_{>0}(z,l_3-1\vert x,0)\, dz\right)\, dx = \nonumber \\
&&\int_0^{+\infty}f(g+x)\left(\int_0^{+\infty}p_{>0}(0,l_1\vert z,0)
p_{>0}(z,l_3-1\vert x,0)\, dz\right)\, dx = \nonumber \\
&&\int_0^{+\infty}f(g+x)p_{>0}(0,l_1+l_3-1\vert x,0)\, dx = \nonumber \\
&&\int_0^{+\infty}f(g+x)p_{>0}(x,l_1+l_3-1\vert 0,0)\, dx = \nonumber \\
&&\int_0^{+\infty}f(g+x)p_{l_1+l_3-1}(x)\, dx =w_2(g,l_1+l_3),
\end{eqnarray}
where we have used $p_{>0}(0,l_1+l_3-1\vert x,0)=p_{>0}(x,l_1+l_3-1\vert 0,0)$, obtained by reversing the direction of time. Consequently, one gets
\begin{equation}\label{eq6.4}
p_n^{br}(g,l_1,l,l_3)=\frac{w_2(g,l)w_2(g,l_1+l_3)}{p(0,n\vert 0,0)}\delta_{l_1+l+l_3,n},
\end{equation}
from which the joint probability distribution of $g$, $l$, and $l^\prime =l_1+l_3$ is readily found to be given by
\begin{equation}\label{eq6.5}
p_n^{br}(g,l,l^\prime)=\frac{l^\prime w_2(g,l)w_2(g,l^\prime)}{p(0,n\vert 0,0)}\delta_{l+l^\prime,n}.
\end{equation}
To proceed we write
\begin{equation}\label{eq6.6}
p_n^{br}(g,l)=\sum_{l^\prime >0}p_n^{br}(g,l,l^\prime)=
\frac{n-l}{p(0,n\vert 0,0)}\mathcal{P}_n(g,l),
\end{equation}
where
\begin{equation}\label{eq6.7}
\mathcal{P}_n(g,l)=\sum_{l^\prime >0}w_2(g,l)w_2(g,l^\prime)\delta_{l+l^\prime,n},
\end{equation}
and we compute the generating function of $\mathcal{P}_n(g,l)$ with respect to $n$ and $l>0$ [we recall that $p_n^{br}(g,-l)=p_n^{br}(g,l)$]. Namely,
\begin{eqnarray}\label{eq6.8}
\sum_{n,l>0}\mathcal{P}_n(g,l)s^l t^n&=&
\left(\sum_{l>0}w_2(g,l)(st)^l\right)\left(\sum_{l^\prime >0}w_2(g,l^\prime)t^{l^\prime}\right)
\nonumber \\
&=&I_1(g,st)I_1(g,t),
\end{eqnarray}
which follows straightforwardly from Eq.\ (\ref{eq2.12}) and the first Eq.\ (\ref{eq2.17}). The large $n$ behavior of $\mathcal{P}_n(g,l)$ is encoded in the large $n$ behavior of its generating function with respect to $l$, $\sum_{l>0}\mathcal{P}_n(g,l)s^l$, which can be extracted by appropriate Tauberian theorems from the behavior of\ (\ref{eq6.8}) as a function of $t$ in the vicinity of its dominant singularity at $t=1$. The latter is obtained without difficulty from Eqs.\ (\ref{eq4.8}) or\ (\ref{eqA3.8}),\ (\ref{eq2.11}), and\ (\ref{eq2.17}). Skipping the details, one finds that for large $n$ the generating function $\sum_{l>0}\mathcal{P}_n(g,l)s^l$ behaves like
\begin{equation}\label{eq6.9}
\sum_{l>0}\mathcal{P}_n(g,l)s^l \sim\frac{\Gamma(1+1/\mu)}{\pi an^{1+1/\mu}}\, 
I_1(g,s)I_2(g) \ \ \ \ (n\rightarrow +\infty).
\end{equation}
On the other hand, one has
\begin{eqnarray}\label{eq6.10}
p(0,n\vert 0,0)&=&\frac{1}{2\pi}\int_{-\infty}^{+\infty}f(k)^n\, dk
= \frac{1}{\pi}\int_0^{+\infty}f(k)^n\, dk \nonumber \\
&=&\frac{1}{\pi\mu an^{1/\mu}}
\int_0^{+\infty}q^{1/\mu -1}f\left(\frac{q^{1/\mu}}{an^{1/\mu}}\right)^n\, dq \nonumber \\
&\sim&\frac{1}{\pi\mu an^{1/\mu}}
\int_0^{+\infty}q^{1/\mu -1}\left(1-\frac{q}{n}\right)^n\, dq \nonumber \\
&\sim&\frac{1}{\pi\mu an^{1/\mu}}
\int_0^{+\infty}q^{1/\mu -1}{\rm e}^{-q}\, dq \nonumber \\
&=&\frac{\Gamma(1+1/\mu)}{\pi an^{1/\mu}} \ \ \ \ (n\rightarrow +\infty),
\end{eqnarray}
which, together with Eq.\ (\ref{eq6.6}), yields
\begin{equation}\label{eq6.11}
p_n^{br}(g,l)\sim\frac{\pi an^{1+1/\mu}}{\Gamma(1+1/\mu)}\mathcal{P}_n(g,l)
 \ \ \ \ (n\rightarrow +\infty),
 \end{equation}
 and, by Eq.\ (\ref{eq6.9}),
\begin{equation}\label{eq6.12}
\sum_{l>0}p_n^{br}(g,l)s^l\sim\frac{\pi an^{1+1/\mu}}{\Gamma(1+1/\mu)}
\sum_{l>0}\mathcal{P}_n(g,l)s^l\sim I_1(g,s)I_2(g)
 \ \ \ \ (n\rightarrow +\infty).
 \end{equation}
 It follows immediately from Eq.\ (\ref{eq6.12}) that $\tilde{p}^{br}(g,s)=\lim_{n\rightarrow +\infty}\sum_{l>0}p_n^{br}(g,l)s^l$ exists and
 \begin{equation}\label{eq6.13}
 \tilde{p}^{br}(g,s)=I_1(g,s)I_2(g).
 \end{equation}
 Finally, by comparing Eqs.\ (\ref{eq2.16}) and\ (\ref{eq6.13}) it can be seen that $\tilde{p}^{br}(g,s)=\tilde{p}(g,s)$, hence the joint distribution of $g$ and $l$ is exactly the same for a bridge and a free-end walk. All the results obtained in Secs.\ \ref{sec3} to\ \ref{sec5} can thus be transposed to the case of a bridge without any modification. It is important to notice that, strictly speaking, this result holds in the limit $n\rightarrow +\infty$ only (infinitely long walk). For a finite long walk ($1\ll n<+\infty$) one expects the subdominant, finite $n$, corrections to be different in the bridge and in free-end walk.
%
%
\section{Numerical simulations}\label{sec7}
In Ref. \cite{MMS} we had already presented some results of numerical computations of $p_{\rm gap}(g)$ and $p_{\rm time}(l)$ for the free-end RW, dealing with different types of jump distributions with various values of $0< \mu \leq 2$. Here, we will focus on the joint distribution $p(g,l)$ for the free-end RW and the marginal distributions $p^{br}_{\rm gap}(g)$ and $p^{br}_{\rm time}(l)$ for the bridge.  
%
%
\subsection{$\bm{p(g,l)}$ for the free-end random walk}
Computing a joint PDF numerically is notoriously challenging. So, instead of $p(g,l)$ we consider the following cumulative distribution,
\begin{equation}\label{p_cumul}
p_>(g,l) = \sum_{m=l}^\infty p(g,m) \;,
\end{equation}
which yields better statistics at a lesser cost. We have computed $p_>(g,l)$ numerically for two different kinds of jump distribution belonging respectively to class A and class C jumps in the classification presented in section\ \ref{sec4}: namely, jump distributions corresponding to L\'evy flights of index $0<\mu <2$ and super-exponential jump distributions.     
\subsubsection{L\'evy flights ($0<\mu<2$)}
It has been shown in Sec.\ \ref{sec4.1.2} that the large $g$ and $l$ behavior of $p(g,l)$ for a L\'evy flight of index $\mu$ takes the scaling form\ (\ref{eq4.1.2.3}). This scaling form of $p(g,l)$ implies the following one for $p_>(g,l)$:
\begin{equation}\label{eq:scaling_cumul}
p_>(g,l) \sim \frac{1}{al^{1 + 1/\mu}} \, G_\mu\left(\frac{g}{a\, l^{1/\mu}}\right)
 \ \ \ \ (l,g\rightarrow +\infty),
\end{equation} 
with
\begin{equation}\label{eq:Gmu}
G_\mu(z) = \frac{1}{z^{1+\mu}} \int_{1/z^{\mu}}^{+\infty} F_{\mu}(y) \, dy.
\end{equation}
The small and large $z$ behaviors of $G_\mu(z)$ are readily obtained from the asymptotic behaviors\ (\ref{eq4.1.2.4}) and\ (\ref{eq4.1.2.6}) of $F_\mu(y)$. One finds,
\begin{equation}\label{asympt_Gmu}
G_\mu(z) \sim
\left\lbrace
\begin{array}{ll}
\mu A_\mu\, z^{-\mu}&(z \to 0), \\
2^{-1}\tilde{C}_\mu\, z^{-1-\mu}&(z \to +\infty),
\end{array}
\right.
\end{equation}
where $A_\mu$ is given by Eq.\ (\ref{eq4.1.2.5}) and $\tilde{C}_\mu$ is the integral\ (\ref{eq4.1.1.14}), with $\alpha =\mu$, the value of which is given in Eq.\ (\ref{eq3.22}). Namely, $A_\mu =\pi^{-1}\Gamma(1+1/\mu)\, \Gamma(1-\mu/2)^{-2}$ and $\tilde{C}_\mu =\mu\, \Gamma(1-\mu/2)^{-2}$. 
\begin{figure}[ht]
\includegraphics[width = 0.6\linewidth]{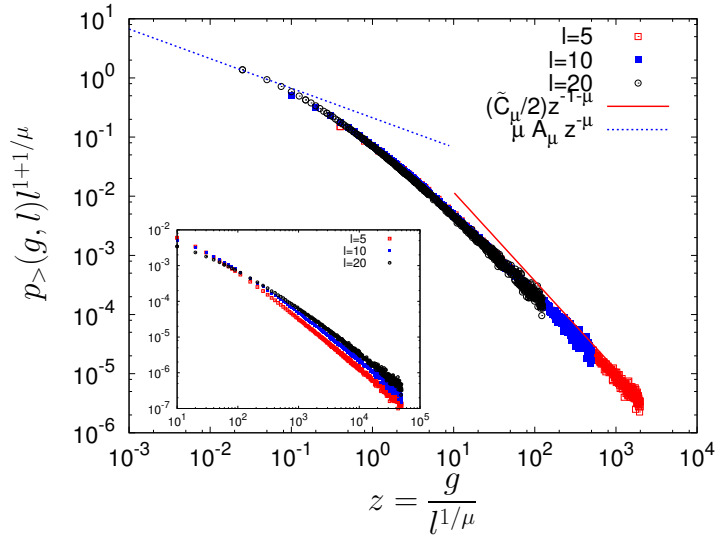}
\caption{Scaled plot of $l^{1+1/\mu} p_>(g,l)$ as a function of the scaling variable $z = g/l^{1/\mu}$ (here $a=1$) for $\mu = 1/2$ and different values of $l=5,\, 10,\, 20$. The lines are guide to eyes, indicating the expected algebraic behaviors in Eq. (\ref{asympt_Gmu}). {\bf Inset:} plot of the same data but without the rescaling. Here $n=100$.}\label{fig_check_scaling}
\end{figure}

In Fig. \ref{fig_check_scaling} we show a plot of $l^{1+1/\mu} p_>(g,l)$ as a function of the scaling variable $z = gl^{-1/\mu}$ (here $a=1$) for different values of $l=5, 10, 20$. We observe that the data for different values of $l$ collapse on a single master curve, which corroborates the scaling form in Eq. (\ref{eq:scaling_cumul}). Note that while this scaling form holds, in principle, only when both $g$ and $l$ are large, numerical simulations show that it works already very well also for moderate values of $g$ or $l$.
%
%
\subsubsection{Super-exponential jump distributions}
We now consider the case of a super-exponential jump distribution,
\begin{equation}\label{super_exp}
f(\eta) = \frac{1}{2 \Gamma(1+1/\alpha)} \exp{\left(- |\eta|^\alpha \right)}\;,
\end{equation}
where $\alpha >1$. It can be checked that $f(\eta)$ in Eq.\ (\ref{super_exp}) satisfies the asymptotic behavior\ (\ref{eq4.3.1}) with $c = \alpha$ and $\delta = \alpha - 1>0$, together with the condition on the function $\theta(x,g)$ below Eq. (\ref{eq4.3.1}). Thus, this case belongs to Class C jumps studied in Section\ \ref{sec4.3} and, as such, should show the concentration phenomenon\ (\ref{eq4.3.8}) which reads, in terms of $p_>(g,l)$,
\begin{equation}\label{ratio}
\frac{p_>(g,1)}{p_>(g,l)}  \underset{g \to \infty} \sim r_l \, g^\delta \;,
\end{equation}
with $l\ge 2$ and where $r_l$ is a $l$-dependent constant, independent of $g$ and unimportant here.
\begin{figure}[h]
\includegraphics[width=0.6\linewidth]{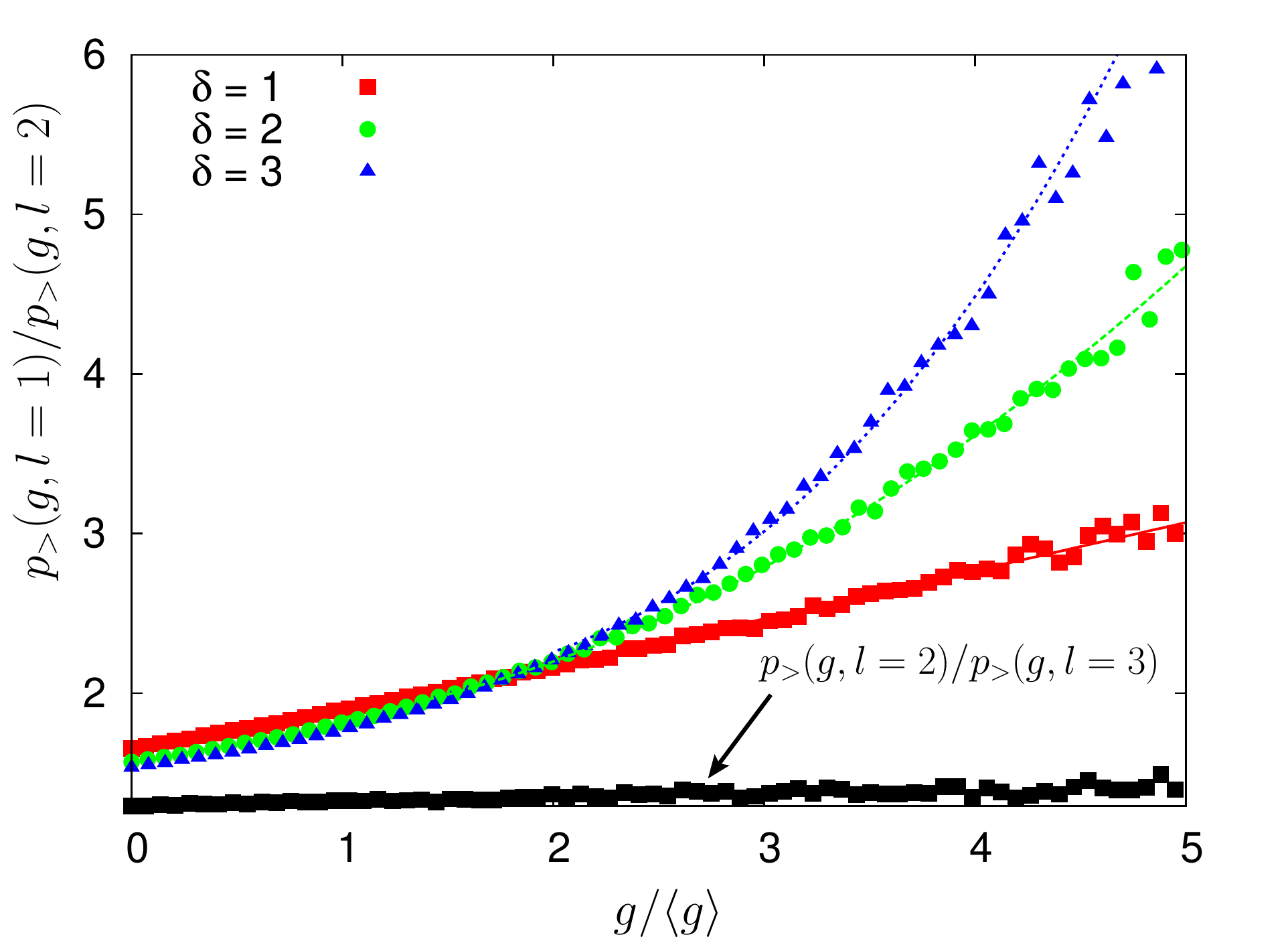}
\caption{Plot of ${p_>(g,1)}/{p_>(g,2)}$ as a function of $g/\langle g \rangle$ from a sample of $10^7$ independent RWs of $n=10^3$ steps for the super-exponential jump distribution\ (\ref{super_exp}) with different values of $\delta = \alpha - 1$. The lines are guide to the eyes, indicating the expected power law growth $\propto g^\delta$ (\ref{ratio}). The lower curve shows ${p_>(g,2)}/{p_>(g,3)}$ for $\delta = 1$, which does not depend on $g/\langle g \rangle$ significantly, in agreement with Eqs.\ (\ref{eq4.3.6}) and\ (\ref{p_cumul}). (Similar results, not shown here, have been obtained for $\delta =2$ and $3$.)}\label{fig_concent}
\end{figure}
Figure \ref{fig_concent} shows a plot of the ratio ${p_>(g,1)}/{p_>(g,2)}$ as a function of $g/\langle g \rangle$ for different values of $\delta = 1, 2$, and $3$ (with $\langle g \rangle = 0.291, 0.261$, and $0.251$, respectively). These data have been obtained by sampling $10^7$ independent RWs of $n=10^3$ steps. For large $g/\langle g \rangle$, numerical results are found to be in a qualitative agreement with the power law behavior $\propto g^\delta$ predicted by Eq. (\ref{ratio}). The difficulty in pinpointing a precise power law behavior in this range of $g$ is due to the poor sampling of large $g$ values for such rapidly decaying jump distributions. On the other hand, a very good agreement with Eq.\ (\ref{ratio}) can already be observed for small to moderate values of $g/\langle g \rangle$, where the sampling is good. Note that these numerical simulations show that the asymptotic expression\ (\ref{ratio}) actually works very well also for moderate values of $g$ (cf. the similar remark at the end of the previous paragraph). This tendency toward the predicted concentration onto $l=\pm 1$ is supported by the numerically computed behavior of the ratio ${p_>(g,2)}/{p_>(g,3)}$ as a function of $g$. In contrast to the growth of ${p_>(g,1)}/{p_>(g,2)}$, we observed no variation of ${p_>(g,2)}/{p_>(g,3)}$ other than numerical noise over the same range $0<g/\langle g \rangle<5$. The lower curve in Fig. \ref{fig_concent} corresponds to $\delta =1$. We have obtained similar results for $\delta =2$ and $3$.
%
%
\subsection{Random bridge}
In general, the numerical simulation of a random bridge, i.e. a RW constrained to start and end at the same position, cannot be ``directly'' constructed for a generic jump distribution $f(\eta)$. There is one exception, however, when the jumps are Gaussian distributed: $f(\eta) = e^{-x^2/(2 \sigma^2)}/\sqrt{2 \pi \sigma^2}$. Indeed, in this case the random variable
\begin{equation}\label{def_RB}
x^{br}_k = x_k - \frac{k}{n}x_n \;,\ \ (0\le k\le n)\;,
\end{equation}
turns out to be equal in law to the position of the walker after the $k$th step of a bridge with $n$ steps and $x^{br}_0 =x^{br}_n =0$ \cite{feller}. In Eq.\ (\ref{def_RB}), $x_k$ is the position of the walker after the $k$th step of the free-end RW\ (\ref{def_RW}) which can be easily computed numerically. Thus, Eq.\ (\ref{def_RB}) gives a very simple way to construct a random bridge from a free-end RW in the case of Gaussian jumps.

The numerical data used in the following have been obtained from\ (\ref{def_RB}) in which the sequence $\lbrace x_k\rbrace$ has been drawn from samples of $10^7$ or $10^6$ independent RWs\ (\ref{def_RW}) with $n=10^4$ and Gaussian jumps with $\sigma = 1$.
\begin{figure}[ht]
\includegraphics[width = \linewidth]{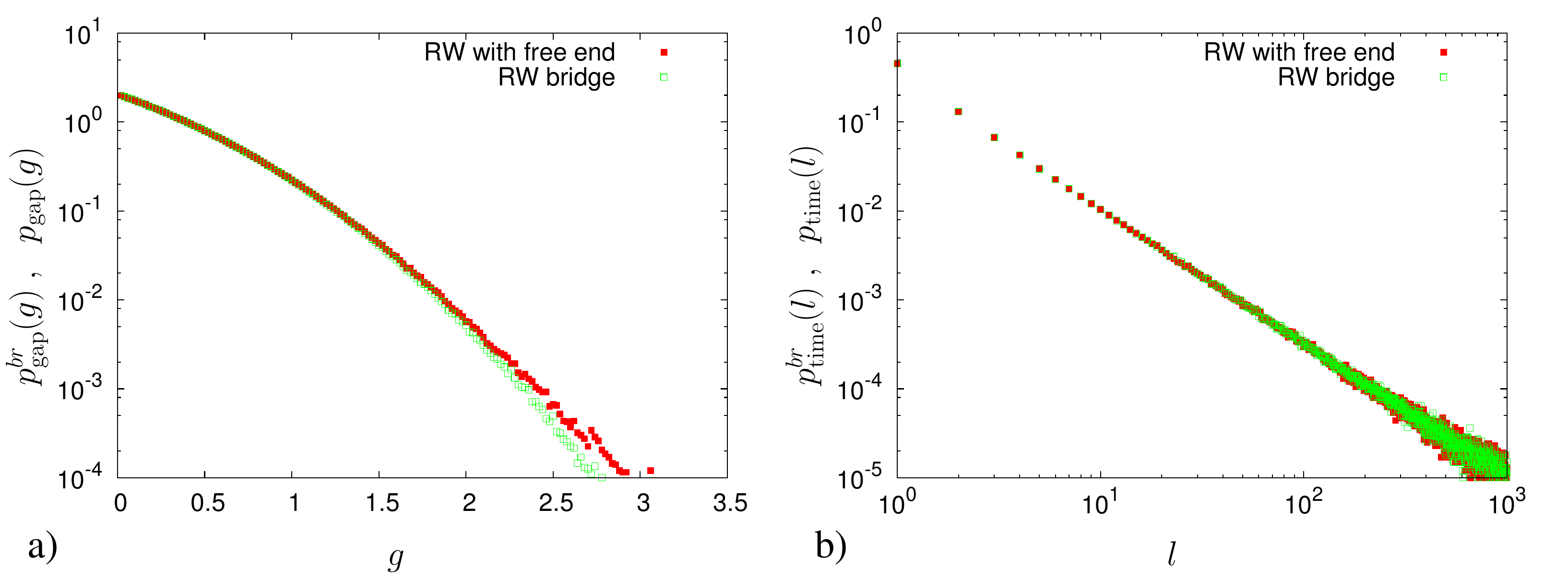}
\caption{{\bf a)} Numerical computation of $p_{\rm gap}(g)$ for a free RW and a RW bridge with Gaussian jumps. {\bf b)} Numerical computation of $p_{\rm time}(l)$ for a free RW and a RW bridge with Gaussian jumps. In both cases, the size of the RW is $n=10^4$ steps.}\label{fig:bridge}
\end{figure}
Figure \ref{fig:bridge} a) shows a plot of the PDF of the gap for the bridge, $p^{br}_{\rm gap}(g)$, and compare it to its counterpart for the free-end RW, $p_{\rm gap}(g)$. Both have been computed numerically from a sample of $=10^7$ RWs. The two PDFs coincide up to $g \sim 2$ while they differ slightly for larger values $g > 2$. This discrepancy between the tails of $p^{br}_{\rm gap}(g)$ and $p_{\rm gap}(g)$ is a finite $n$ effect. We have checked that increasing the value of $n$ does extend the agreement between the two PDFs to larger and larger values of $g$, as expected. Figure \ref{fig:bridge} b) shows a plot of the PDF of the time between the first two maxima for the bridge, $p^{br}_{\rm gap}(l)$, and compare it to its counterpart for the free-end RW, $p_{\rm time}(l)$. Both have been computed numerically from a sample of $=10^6$ RWs. The agreement between the two PDFs is very good. These numerical results bear out the conclusion of Sec.\ \ref{sec6} according to which the stationary PDFs for the bridge and the free-end RW coincide.  

While we have studied here the case of Gaussian jumps, the case of arbitrary jump distribution could also be studied numerically using Monte-Carlo methods for constrained RWs, as done for instance for L\'evy flights in Ref. \cite{SM10}. 
%
%
\section{Summary and perspectives}\label{sec8}
To summarize, we have performed a detailed analytical study of the statistics of the gap, $G_n$, and the time interval, $L_n$, between the two highest positions of a RW of $n$ steps (see Fig.\ \ref{Fig_markov}) in the limit where $n \to \infty$. The results we have obtained are quite general as we have addressed the question for the wide class of a symmetric, bounded, and (piecewise) continuous jump distribution, including in particular the ubiquitous case of L\'evy flights of index $\mu$, with $0<\mu< 2$. Using first-passage techniques, including the powerful Hopf-Ivanov formula, we have first shown that the joint PDF of $G_n$ and $L_n$, $p_n(g,l)$, converges to a stationary PDF, $p_n(g,l) \to p(g,l)$, as $n \to \infty$. We have then obtained an explicit expression of the GF of the stationary joint PDF $p(g,l)$ w.r.t. $l$, given in Eqs. (\ref{eq2.16}) and (\ref{eq2.17}), from which most our results have been derived. We have found a quite rich landscape of behaviors of $p(g,l)$ in the $(l,g)$ plane, summarized in Figs.\ \ref{fig2} to\ \ref{fig4}, depending on the tail of the jump distribution: algebraic, exponential, or super-exponential. For instance, in the first case (algebraic tail) and in the scaling regime $g,\, l \gg 1$ with fixed $l g^{-\mu}$, we have shown that $p(g,l)$ takes a scaling form, Eq.\ (\ref{eq4.1.1.8}), while in the case of a super-exponential tail, there is an unexpected concentration of $p(g,l)$ onto $l=\pm 1$ as $g$ gets large. We have computed the stationary marginal distributions of the gap $G_n$, $p_{\rm gap}(g)$, and of the time $L_n$, $p_{\rm time}(l)$, from the stationary joint PDF $p(g,l)$. Their asymptotic behaviors, which depend on the L\'evy index $\mu$, are summarized in Eqs. (\ref{eq:result_p_of_g}) and (\ref{eq:result_p_of_l}). The latter reveals an unexpected freezing phenomenon of the exponent characterizing the algebraic tail of $p_{\rm time}(l)$ as $\mu$ decreases past the value $\mu_c = 1$. As a consequence, the first moment of $p_{\rm time}(l)$ is never defined.

We have also considered the joint PDF $p_n^{br}(g,l)$ of $G_n$ and $L_n$ for a bridge, i.e. a random walk constrained to start and end at the origin after $n$ steps. In this case, and using the same approach as for the free-end RW, we have found that $p_n^{br}(g,l)$ converges to the same stationary distribution as for the free-end RW, $p_n^{br}(g,l) \to p(g,l)$, when $n \to \infty$. Note that this interesting result was not obvious at all, {\it a priori}. Hence all the results derived for the free-end RW in the large $n$ limit also hold for the bridge (to within small corrections that go to zero as $n \to \infty$).

The problem considered in the present work constitutes one rare instance of strongly correlated random variables where gap statistics can be computed exactly. There are various directions along which this study could be pushed further. A first natural question concerns the extension to the case of a discrete jump distribution, which is not covered by the present analysis. Note that in the simplest case where $\eta_i = \pm 1$, the problem should be suitably adapted, as the gap between the two highest positions is always one, $G_n = 1$, (and $\vert L_n\vert = 1$ as well). In this case, a related more relevant question to address may be the statistics of the local maxima. 

Another natural extension of the present work concerns the study of higher order gaps $d_{k,n}$, between the $k$-th and $(k+1)$-th maximum of the RW after $n$ steps (here we have studied the case $k=1$, $G_n = d_{1,n}$). The higher order gap statistics for $\mu=2$ was studied in Ref.~\cite{SM12}, where a quite interesting behavior was found in the limit of large $k$. In this case $\mu =2$, the analysis of the statistics of the time elapsed between the $k$-th and $(k+1)$-th maximum is still an open question. Finally, the study of higher order gap statistics for L\'evy flights, $0<\mu < 2$, remains a challenging problem. 
\acknowledgments{SNM and GS acknowledge support by ANR grant 2011-BS04-013-01 WALKMAT and in part by the Indo-French 
Centre for the Promotion of Advanced Research under Project 4604-3. G. S. also acknowledges support from Labex PALM (Project RANDMAT).}
%
%
%
%
\appendix
\section{The Hopf-Ivanov formula}\label{app1}
In this appendix we derive the so-called Hopf-Ivanov formula, explain how to get Eqs.~(\ref{eq2.6}) and\ (\ref{eq2.7}) from it, and prove the expression\ (\ref{eq2.8}) for $\phi(\lambda ,s)$.

Consider a positive walk starting from $x_1\ge 0$ at time $0$ and arriving at $x_2\ge 0$ at time $n$. Assume that the walker reaches the minimum $y\ge 0$ of the walk at time $n_1$ and split the walk into two successive parts: from $x_1$ at time $0$ to $y$ at time $n_1$, then from $y$ at time $n_1$ to $x_2$ at time $n$. By taking $y$ as a new origin in both parts, reversing time direction in the first one, and taking into account all the possible values of $y$ and $n_1$, one has
\begin{eqnarray}\label{eqA1.1}
&&p_{>0}(x_2,n\vert x_1,0)=\sum_{n_1=1-\sigma}^{n-\sigma}\int_0^{\min(x_1,x_2)}
p_{>0}(x_1-y,n_1\vert 0,0)p_{>0}(x_2-y,n-n_1\vert 0,0)\, dy, \nonumber \\
&&=\sum_{n_1\ge 1-\sigma}\sum_{n_2\ge\sigma}\left\lbrack\int_0^{\min(x_1,x_2)}
p_{>0}(x_1-y,n_1\vert 0,0)p_{>0}(x_2-y,n_2\vert 0,0)\, dy\right\rbrack\, \delta_{n_1+n_2,n},
\end{eqnarray}
where $n_2=n-n_1$ and $\sigma =1$ (resp. $0$) if $\min(x_1,x_2)=x_1$ (reps. $x_2$). The sums on the right-hand side of\ (\ref{eqA1.1}) can actually be continued to $n_1,n_2\ge 0$, (the difference corresponds to a term proportional to $\delta\lbrack y-\max(x_1,x_2)\rbrack$ in the integral over $y$ the contribution of which is zero), and\ (\ref{eqA1.1}) reads
\begin{equation}\label{eqA1.2}
p_{>0}(x_2,n\vert x_1,0)=
\sum_{n_1\ge 0}\sum_{n_2\ge 0}\left\lbrack\int_0^{\min(x_1,x_2)}
p_{>0}(x_1-y,n_1\vert 0,0)p_{>0}(x_2-y,n_2\vert 0,0)\, dy\right\rbrack\, \delta_{n_1+n_2,n}.
\end{equation}
Let
\begin{equation}\label{eqA1.3}
G_{>0}(x_2,x_1,s)=\sum_{n\ge 0}p_{>0}(x_2,n\vert x_1,0)s^n.
\end{equation}
From\ (\ref{eqA1.2}) and\ (\ref{eqA1.3}) one gets
\begin{equation*}
G_{>0}(x_2,x_1,s)=\int_0^{\min(x_1,x_2)}
\left(\sum_{n_1\ge 0}p_{>0}(x_1-y,n_1\vert 0,0)s^{n_1}\right)
\left(\sum_{n_2\ge 0}p_{>0}(x_2-y,n_2\vert 0,0)s^{n_2}\right)\, dy.
\end{equation*}
According to $u(x,s)=\sum_{n\ge 0}p_{>0}(x,n\vert 0,0)s^n$ [see Eqs.\ (\ref{eq2.7}) and\ (\ref{eq2.11})], this can be rewritten as
\begin{equation}\label{eqA1.4}
G_{>0}(x_2,x_1,s)=\int_0^{\min(x_1,x_2)}
u(x_1-y,s)u(x_2-y,s)\, dy \;,
\end{equation}
which coincides with Eq. (9) of\ \cite{Iv}. Laplace transforming\ (\ref{eqA1.4}) with respect to both $x_1$ and $x_2$, one finds
\begin{eqnarray}\label{eqA1.5}
&&\int_0^{+\infty}\int_0^{+\infty}
G_{>0}(x_2,x_1,s){\rm e}^{-\lambda_1 x_1-\lambda_2 x_2}dx_1\, dx_2 \nonumber \\
&&=\int_0^{+\infty}\int_0^{+\infty}\left(\int_0^{\min(x_1,x_2)}
u(x_1-y,s)u(x_2-y,s)\, dy\right){\rm e}^{-\lambda_1 x_1-\lambda_2 x_2}dx_1\, dx_2 \nonumber \\
&&=\int_0^{+\infty}\left(\int_y^{+\infty}u(x_1-y,s){\rm e}^{-\lambda_1 x_1}dx_1\right)
\left(\int_y^{+\infty}u(x_2-y,s){\rm e}^{-\lambda_2 x_2}dx_2\right)\, dy \nonumber \\
&&=\frac{1}{\lambda_1+\lambda_2}
\left(\int_0^{+\infty}u(z_1,s){\rm e}^{-\lambda_1 z_1}dz_1\right)
\left(\int_0^{+\infty}u(z_2,s){\rm e}^{-\lambda_2 z_2}dz_2\right),
\end{eqnarray}
where we have made the change of variables $z_1=x_1-y$ and $z_2=x_2-y$. Finally, using the first Eq.\ (\ref{eq2.11}) on the right-hand side and replacing $G_{>0}(x_2,x_1,s)$ with its definition\ (\ref{eqA1.3}) on the left-hand side, one obtains the Hopf-Ivanov formula
\begin{equation}\label{eqA1.6}
\int_0^{+\infty}\int_0^{+\infty}
\sum_{n\ge 0}p_{>0}(x_2,n\vert x_1,0)s^n\, 
{\rm e}^{-\lambda_1 x_1-\lambda_2 x_2}dx_1\, dx_2=
\frac{\phi(\lambda_1 ,s)\phi(\lambda_2 ,s)}{\lambda_1+\lambda_2}.
\end{equation}
The Pollaczek-Spitzer formula\ (\ref{eq2.6}) is the Hopf-Ivanov formula\ (\ref{eqA1.6}) for $\lambda_1 =\lambda$ and $\lambda_2 =0$. Note that $\phi(0,s)=1/\sqrt{1-s}$ as readily seen from Eq.\ (\ref{eq2.8}) in the limit $\lambda\rightarrow 0$, (see also e.g.\ \cite{AS2005}). As for Equation\ (\ref{eq2.7}), it is the limit $\lambda_1\rightarrow +\infty$ of\ (\ref{eqA1.6}) with $\lambda_2=\lambda$. [Make the change of variable $x_1=\overline{x}_1/\lambda_1$ and use $\phi(+\infty ,s)=1$].

It remains to prove the expression\ (\ref{eq2.8}) of $\phi(\lambda ,s)$. We will follow the same line as Ivanov in\ \cite{Iv}. From
\begin{equation*}
p_{>0}(x_2,n\vert x_1,0)=\int_0^{+\infty}f(x_2-y)p_{>0}(y,n-1\vert x_1,0)\, dy,
\end{equation*}
the definition\ (\ref{eqA1.3}), and $p_{>0}(x_2,0\vert x_1,0)=\delta(x_2-x_1)$ one gets the integral equation
\begin{equation}\label{eqA1.7}
G_{>0}(x_2,x_1,s)-s\int_0^{+\infty}f(x_2-y)G_{>0}(y,x_1,s)\, dy =\delta(x_2-x_1),
\end{equation}
which for $x_1=0$ and $x_2=x$ reads
\begin{equation}\label{eqA1.8}
u(x,s)-s\int_0^{+\infty}f(x-y)u(y,s)\, dy =\delta(x) \;,
\end{equation}
where we have used $G_{>0}(x,0,s)=\sum_{n\ge 0}p_{>0}(x,n\vert 0,0)s^n=u(x,s)$. Here and in the following, $\delta(x)$ is to be understood as $\delta(x-0^+)$. Deriving\ (\ref{eqA1.8}) with respect to $s$ and using\ (\ref{eqA1.8}) in the result, one has
\begin{equation}\label{eqA1.9}
\frac{\partial u(x,s)}{\partial s}
-s\int_0^{+\infty}f(x-y)\frac{\partial u(x,s)}{\partial s}\, dy =
\frac{u(x,s)-\delta(x)}{s}.
\end{equation}
From\ (\ref{eqA1.9}) and\ (\ref{eqA1.7}) it follows
\begin{equation}\label{eqA1.10}
s\frac{\partial u(x,s)}{\partial s}=\int_0^{+\infty}G_{>0}(x,y,s)
\lbrack u(y,s)-\delta(y)\rbrack\, dy,
\end{equation}
and, by Laplace transforming\ (\ref{eqA1.10}) with respect to $x$,
\begin{eqnarray}\label{eqA1.11}
s\frac{\partial \phi(\lambda ,s)}{\partial s}&=&\int_0^{+\infty}\tilde{G}_{>0}(\lambda ,y,s)
\lbrack u(y,s)-\delta(y)\rbrack\, dy \nonumber \\
&=&\int_0^{+\infty}\tilde{G}_{>0}(\lambda ,y,s)u(y,s)\, dy -\tilde{G}_{>0}(\lambda ,0,s).
\end{eqnarray}
Here, $\lambda$ is the Laplace variable and
\begin{eqnarray}\label{eqA1.12}
\tilde{G}_{>0}(\lambda ,y,s)&=&\int_0^{+\infty}G_{>0}(x,y,s){\rm e}^{-\lambda x}dx \nonumber \\
&=&\int_0^{+\infty}\int_0^{\min(x,y)}
u(x-z,s)u(y-z,s){\rm e}^{-\lambda x}\, dz\, dx \nonumber \\
&=&\int_0^{y}u(y-z,s)\int_z^{+\infty}u(x-z,s){\rm e}^{-\lambda x}\, dx\, dz \nonumber \\
&=&\phi(\lambda ,s)\int_0^{y}u(y-z,s){\rm e}^{-\lambda z}dz,
\end{eqnarray}
where we have used\ (\ref{eqA1.4}) and\ (\ref{eq2.11}). Note that Equation\ (\ref{eqA1.12}) reduces to $\tilde{G}_{>0}(\lambda ,0,s)=\phi(\lambda ,s)$ for $y=0$ which can also be obtained directly from Eq.\ (\ref{eq2.11}) and $u(x,s)=G_{>0}(x,0,s)$. Substituting for $\tilde{G}_{>0}(\lambda ,y,s)$ on the right-hand side of\ (\ref{eqA1.11}) its expression from\ (\ref{eqA1.12}), one finds
\begin{eqnarray}\label{eqA1.13}
s\frac{\partial \phi(\lambda ,s)}{\partial s}&=&\phi(\lambda ,s)\left\lbrack
\int_0^{+\infty}\int_0^y u(y-z,s)u(y,s){\rm e}^{-\lambda z}\, dz\, dy -1\right\rbrack \nonumber \\
&=&\phi(\lambda ,s)\left\lbrack\int_0^{+\infty}\left(\int_z^{+\infty}
u(y-z,s)u(y,s)\, dy\right) {\rm e}^{-\lambda z}dz -1\right\rbrack \nonumber \\
&=&\phi(\lambda ,s)\left\lbrack\int_0^{+\infty}\left(\int_0^{+\infty}u(t,s)u(t+z,s)\, dt\right)
{\rm e}^{-\lambda z}dz -1\right\rbrack \nonumber \\
&=&\phi(\lambda ,s)\int_0^{+\infty}\left(\int_0^{+\infty}u(t,s)u(t+z,s)\, dt -\delta(z)\right)
{\rm e}^{-\lambda z}dz,
\end{eqnarray}
where $t=y-z$. To proceed we take $x_1=X$ and $x_2=X+x$, $x\ge 0$, in Eq.\ (\ref{eqA1.4}). Then, we make the change of variable $y=X-t$, and let $X\rightarrow +\infty$. Using $\lim_{X\rightarrow +\infty}G_{>0}(X+x,X,s)=G_{>-\infty}(x,0,s)$, one finds
\begin{equation}\label{eqA1.14}
G_{>-\infty}(x,0,s)=\int_0^{+\infty}
u(t,s)u(t+x,s)\, dt,
\end{equation}
which coincides with Eq. (13) of\ \cite{Iv}. Thus, Eq.\ (\ref{eqA1.13}) reduces to
\begin{equation}\label{eqA1.15}
s\frac{\partial\ln\phi(\lambda ,s)}{\partial s}=\tilde{\Xi}(\lambda ,s),
\end{equation}
where $\tilde{\Xi}(\lambda ,s)$ is the Laplace transform of $\Xi(x,s)=G_{>-\infty}(x,0,s)-\delta(x)$ with respect to $x$. The equation for $G_{>-\infty}(x,0,s)$ is similar to\ (\ref{eqA1.8}) with a full-space integration, which is readily solved by Fourier transform. One gets
\begin{equation}\label{eqA1.16}
G_{>-\infty}(x,0,s)=\frac{1}{2\pi}\int_{-\infty}^{+\infty}\frac{\exp(-ikx)}{1-s\hat{f}(k)}\, dk,
\end{equation}
hence
\begin{equation}\label{eqA1.17}
\Xi(x,s)=\frac{1}{2\pi}\int_{-\infty}^{+\infty}\frac{s\hat{f}(k)}{1-s\hat{f}(k)}
{\rm e}^{-ikx}\, dk,
\end{equation}
and, by Fubini's theorem,
\begin{eqnarray}\label{eqA1.18}
\tilde{\Xi}(\lambda ,s)&=&\frac{1}{2\pi}\int_0^{+\infty}{\rm e}^{-\lambda x}\left(\int_{-\infty}^{+\infty}
\frac{s\hat{f}(k)}{1-s\hat{f}(k)}{\rm e}^{-ikx}\, dk\right) dx \nonumber \\
&=&\frac{1}{2\pi}\int_{-\infty}^{+\infty}\frac{s\hat{f}(k)}{1-s\hat{f}(k)}
\left(\int_0^{+\infty}{\rm e}^{-(\lambda+ik)x}\, dx\right) dk \nonumber \\
&=&\frac{1}{\pi}\int_0^{+\infty}
\frac{\lambda}{k^2+\lambda^2}\left\lbrack\frac{s\hat{f}(k)}{1-s\hat{f}(k)}\right\rbrack dk.
\end{eqnarray}
Substituting\ (\ref{eqA1.18}) for $\tilde{\Xi}(\lambda ,s)$ on the right-hand side of\ (\ref{eqA1.15}) yields
\begin{equation}\label{eqA1.19}
\frac{\partial\ln\phi(\lambda ,s)}{\partial s}=\frac{1}{\pi}\int_0^{+\infty}
\frac{\lambda}{k^2+\lambda^2}\left\lbrack\frac{\hat{f}(k)}{1-s\hat{f}(k)}\right\rbrack dk,
\end{equation}
Finally, integrating\ (\ref{eqA1.19}) with $\phi(\lambda ,0)=1$ [which follows from Eq.\ (\ref{eq2.11}) and $u(x,0)=\delta(x)$], one obtains
\begin{equation}\label{eqA1.20}
\ln\phi(\lambda ,s)=-\frac{\lambda}{\pi}
\int_0^{+\infty}\frac{\ln\lbrack 1-s\hat{f}(k)\rbrack}{k^2+\lambda^2}\, dk,
\end{equation}
which completes the proof of Eq.\ (\ref{eq2.8}).
%
%
\section{From the small $\bm{k}$ behavior of $\bm{\hat{f}(k)}$ to the large $\bm{\eta}$ behavior of $\bm{f(\eta)}$ for $\bm{0<\mu <2}$}\label{app2}
In this appendix we recall how the large $\eta$ behavior of $f(\eta)$ can be deduced from the small $k$ behavior of $\hat{f}(k)$ when $0<\mu <2$. Since $f(\eta)$ is real and symmetric, $\hat{f}(k)$ is also real and symmetric and one has
\begin{equation}\label{eqA2.1}
f(\eta)=\frac{1}{\pi}{\rm Re}\int_{0}^{+\infty}\hat{f}(k){\rm e}^{-ik\eta}\, dk
=\frac{1}{\pi \eta}{\rm Re}\int_{0}^{+\infty}\hat{f}\left(\frac{q}{\eta}\right){\rm e}^{-iq}\, dq,
\end{equation}
where we have made the change of variable $q=k\eta$.

Consider first $0<\mu <1$. Integrating by parts once in\ (\ref{eqA2.1}) and using $\hat{f}(k)\sim 1-(ak)^\mu$, ($k\rightarrow 0$), one obtains
\begin{eqnarray}\label{eqA2.2}
f(\eta)&=&\frac{1}{\pi \eta^2}{\rm Re}\int_{0}^{+\infty}\hat{f}^\prime\left(\frac{q}{\eta}\right)
{\rm e}^{-iq}\, \frac{dq}{i} \nonumber \\
&\sim&\frac{1}{\pi \eta^{\mu +1}}{\rm Re}\ i\mu a^\mu\int_0^{+\infty}q^{\mu -1}
{\rm e}^{-iq}\, dq \nonumber \\
&=&\sin\left(\frac{\pi\mu}{2}\right)\Gamma(\mu +1)\frac{a^\mu}{\pi \eta^{\mu +1}}
 \ \ \ \ (\eta\rightarrow +\infty),
 \end{eqnarray}
where the integral over $q$ in the second line is equal to $\exp(-i\pi\mu/2)\Gamma(\mu)$.
 
For $1\le\mu <2$ we integrate by parts twice in\ (\ref{eqA2.1}), which gives
\begin{equation}\label{eqA2.3}
f(\eta)\sim\lim_{q\rightarrow 0}\frac{\mu a^\mu q^{\mu -1}}{\pi \eta^{\mu +1}}+
\frac{\mu(\mu -1)a^\mu}{\pi \eta^{\mu +1}}
{\rm Re}\int_0^{+\infty}q^{\mu -2}{\rm e}^{-iq}\, dq
 \ \ \ \ (\eta\rightarrow +\infty).
\end{equation}
If $\mu =1$ the second term on the right-hand side of\ (\ref{eqA2.3}) is zero and one has $f(\eta)\sim a/(\pi \eta^2)$, $(\eta\rightarrow +\infty$). On the other hand, if $1<\mu <2$ it is the first term on the right-hand side of\ (\ref{eqA2.3}) which is zero and one gets
\begin{eqnarray}\label{eqA2.4}
f(\eta)&\sim&\frac{\mu(\mu -1)a^\mu}{\pi \eta^{\mu +1}}
{\rm Re}\int_0^{+\infty}q^{\mu -2}{\rm e}^{-iq}\, dq \nonumber \\
&=&\sin\left(\frac{\pi\mu}{2}\right)\Gamma(\mu +1)\frac{a^\mu}{\pi \eta^{\mu +1}}
 \ \ \ \ (\eta\rightarrow +\infty),
 \end{eqnarray}
where the integral over $q$ in the first line is equal to $i\exp(-i\pi\mu/2)\Gamma(\mu -1)$. Thus, the large $\eta$ behavior of $f(\eta)$ for $0<\mu <2$ is always given by the expression
\begin{equation}\label{eqA2.5}
f(\eta)\sim\sin\left(\frac{\pi\mu}{2}\right)\Gamma(\mu +1)\frac{a^\mu}{\pi \eta^{\mu +1}}
 \ \ \ \ (\eta\rightarrow +\infty).
\end{equation}
%
%
\section{Large $\bm{l}$ behavior of $\bm{p(g,l)}$ for integral $\bm{1/\mu}$}\label{app3}
Our starting point is similar to the one at the beginning of Sec.\ \ref{sec4}. Namely, we expand the logarithm on the right-hand side of\ (\ref{eq4.1}) in power series of $(1-s)$ up to order $1/\mu -1$, and we write
\begin{equation}\label{eqA3.1}
\int_0^{+\infty}\frac{\lambda\, \ln\lbrack 1+(1-s)\hat{F}(k)\rbrack}{k^2+\lambda^2}\, dk
= \sum_{n=1}^{1/\mu -1}\frac{(-1)^{n+1}}{n}\beta_n(\lambda)(1-s)^n+R(\lambda ,s),
\end{equation}
where $\beta_n(\lambda)$ is defined by\ (\ref{eq4.7}) and $R(\lambda ,s)$ is the remainder,
\begin{equation}\label{eqA3.2}
R(\lambda ,s)=\int_0^{+\infty}\frac{\lambda}{k^2+\lambda^2}\left\lbrace
\ln\lbrack 1+(1-s)\hat{F}(k)\rbrack +
\sum_{n=1}^{1/\mu -1}\frac{(-1)^n}{n}(1-s)^n\hat{F}(k)^n\right\rbrace\, dk.
\end{equation}
Since $n\le 1/\mu -1$ and $\hat{F}(k)\sim k^{-\mu}$ as $k\rightarrow 0$, the existence of $\beta_n(\lambda)$ and of the integral over $k$ in\ (\ref{eqA3.2}) is ensured. Now, we cannot continue straightforwardly like in Sec.\ \ref{sec4} because\ (\ref{eq4.5}) does not exist if $1/\mu$ is an integer. To proceed we pick a given $k_c>0$ arbitrarily small and split the integral\ (\ref{eqA3.2}) into $R(\lambda ,s)=R^{-}(\lambda ,s)+R^{+}(\lambda ,s)$, with
\begin{equation}\label{eqA3.3}
R^{-}(\lambda ,s)=\int_0^{k_c}\frac{\lambda}{k^2+\lambda^2}\left\lbrace
\ln\lbrack 1+(1-s)\hat{F}(k)\rbrack +
\sum_{n=1}^{1/\mu -1}\frac{(-1)^n}{n}(1-s)^n\hat{F}(k)^n\right\rbrace\, dk,
\end{equation}
and
\begin{equation}\label{eqA3.4}
R^{+}(\lambda ,s)=\int_{k_c}^{+\infty}\frac{\lambda}{k^2+\lambda^2}\left\lbrace
\ln\lbrack 1+(1-s)\hat{F}(k)\rbrack +
\sum_{n=1}^{1/\mu -1}\frac{(-1)^n}{n}(1-s)^n\hat{F}(k)^n\right\rbrace\, dk.
\end{equation}
Taking $k_c$ small enough, we can replace $\hat{F}(k)$ with $(ak)^{-\mu}$ and ignore $k^2$ compared to $\lambda^2$ in $R^{-}(\lambda ,s)$. Making then the change of variable $k=(1-s)^{1/\mu}q/a$, one gets
\begin{equation}\label{eqA3.5}
R^{-}(\lambda ,s) =\frac{(1-s)^{1/\mu}}{a\lambda}
\int_0^{ak_c/(1-s)^{1/\mu}}\left\lbrack\ln\left(1+\frac{1}{q^\mu}\right)+\sum_{n=1}^{1/\mu -1}
\frac{(-1)^n}{nq^{n\mu}}\right\rbrack\, dq.
\end{equation}
In the limit $s\rightarrow 1$, the integral over $q$ in\ (\ref{eqA3.5}) diverges logarithmically and $R^{-}(\lambda ,s)$ behaves like
\begin{equation}\label{eqA3.6}
R^{-}(\lambda ,s) =\frac{(-1)^{1+1/\mu}}{a\lambda}(1-s)^{1/\mu}\ln\left(\frac{1}{1-s}\right)
+O(1-s)^{1/\mu}.
\end{equation}
On the other hand, it is readily seen that $R^{+}(\lambda ,s)$ is analytic at $s=1$ with $R^{+}(\lambda ,s)=O(1-s)^{1/\mu}$. Thus, in the vicinity of $s=1$\ (\ref{eqA3.1}) reads
\begin{eqnarray}\label{eqA3.7}
\int_0^{+\infty}\frac{\lambda\, \ln\lbrack 1+(1-s)\hat{F}(k)\rbrack}{k^2+\lambda^2}\, dk
&=& \sum_{n=1}^{1/\mu -1}\frac{(-1)^{n+1}}{n}\beta_n(\lambda)(1-s)^n \\
&-&\frac{(-1)^{1/\mu}}{a\lambda}(1-s)^{1/\mu}\ln\left(\frac{1}{1-s}\right)
+O(1-s)^{1/\mu}, \nonumber
\end{eqnarray}
which replaces Eq.\ (\ref{eq4.6}) when $1/\mu$ is an integer. Now, it remains to follow the same line as below Eq.\ (\ref{eq4.6}). One finds
\begin{eqnarray}\label{eqA3.8}
\phi(\lambda ,s)&=&\phi(\lambda ,1)\left\lbrack 1+
\sum_{n=1}^{1/\mu -1}\gamma_n(\lambda)(1-s)^n\right. \\
&+&\left.\frac{(-1)^{1/\mu}}{\pi a}\frac{(1-s)^{1/\mu}}{\lambda}\ln\left(\frac{1}{1-s}\right)
+O(1-s)^{1/\mu}\right\rbrack, \nonumber
\end{eqnarray}
and
\begin{eqnarray}\label{eqA3.9}
\tilde{p}(g,s)&=&\tilde{p}(g,1) +I_2(g)\sum_{n=1}^{1/\mu -1}J_n(g)(1-s)^n \\
&+&\frac{(-1)^{1/\mu}}{\pi a} I_2(g)^2 (1-s)^{1/\mu}\ln\left(\frac{1}{1-s}\right)
+O(1-s)^{1/\mu} \nonumber,
\end{eqnarray}
which replace Eqs.\ (\ref{eq4.8}) and\ (\ref{eq4.11}), respectively. The large $l$ behavior of $p(g,l)$ at fixed $g$ is determined by the singular term $(1-s)^{1/\mu}\ln\lbrack 1/(1-s)\rbrack$ on the right-hand side of\ (\ref{eqA3.9}) through the appropriate Tauberian theorem. According to\ \cite{FS}, [Chap. VI, Eq. (26)], a factor $(1-s)^{1/\mu}\ln\lbrack 1/(1-s)\rbrack$,\ ($s\rightarrow 1$), in $\tilde{p}(g,s)$ translates into a factor $(-1)^{1/\mu}\Gamma(1+1/\mu)/l^{1+1/\mu}$,\ ($l\rightarrow +\infty$), in $p(g,l)$ and one finally obtains
\begin{equation}\label{eqA3.10}
p(g,l)\sim\frac{\Gamma(1+1/\mu)}{\pi a}\, \frac{I_2(g)^2}{l^{1+1/\mu}} \ \ \ \ (l\rightarrow +\infty),
\end{equation}
which is nothing but Eq.\ (\ref{eq4.13}) with integral $1/\mu$.
%
%
\section{Scaling form of $\bm{p(g,l)}$ for jump distributions with an algebraic tail}\label{app4}
In this appendix we derive the scaling form of $p(g,l)$ given in Sec.\ \ref{sec4.1.1}. From Eq.\ (\ref{eq2.16}) one has
\begin{equation}\label{eqA4.1}
p(g,l)=\frac{1}{2i\pi}\oint\frac{\tilde{p}(g,s)}{s^{l+1}}\, ds=
\frac{I_2(g)}{2i\pi}\oint\frac{I_1(g,s)}{s^{l+1}}\, ds,
\end{equation}
with
\begin{equation}\label{eqA4.2}
I_1(g,s)=s\int_0^{+\infty}u(x,s)f(g+x)\, dx=
sg\int_0^{+\infty}u(g\overline{x},s)f\lbrack g(1+\overline{x})\rbrack\, d\overline{x},
\end{equation}
where $\overline{x}=x/g$, and, Laplace inverting the first Eq.\ (\ref{eq2.11}),
\begin{equation}\label{eqA4.3}
u(g\overline{x},s)=\frac{1}{2i\pi}\int_{\cal{L}}\phi(\lambda ,s)
\, {\rm e}^{\lambda g\overline{x}}d\lambda =
\frac{1}{2i\pi g}\int_{\cal{L}}\phi\left(\frac{\overline{\lambda}}{g} ,s\right)
\, {\rm e}^{\overline{\lambda} \overline{x}}d\overline{\lambda},
\end{equation}
where $\overline{\lambda}=g\lambda$ and $\cal{L}$ is a Bromwich contour. From Eq.\ (\ref{eq4.1}) one gets
\begin{equation}\label{eqA4.4}
\phi\left(\frac{\overline{\lambda}}{g} ,s\right)
=\phi\left(\frac{\overline{\lambda}}{g} ,1\right)\exp\left(-\frac{\overline{\lambda}}{\pi}
\int_0^{+\infty}\frac{\ln\lbrack 1+(1-s)\hat{F}(q/g)\rbrack}
{q^2+\overline{\lambda}^2}\, dq\right) ,
\end{equation}
where $q=gk$ and, by letting $g\rightarrow +\infty$,
\begin{eqnarray}\label{eqA4.5}
&&\phi\left(\frac{\overline{\lambda}}{g} ,s\right)
\sim\phi\left(\frac{\overline{\lambda}}{g} ,1\right)\exp\left(-\frac{\overline{\lambda}}{\pi}
\int_0^{+\infty}\frac{1}{q^2+\overline{\lambda}^2}\, 
\ln\left\lbrack 1+\frac{(1-s)}{q^\mu}\left(\frac{g}{a}\right)^\mu\right\rbrack\, dq\right) \nonumber \\
&&\sim\left(\frac{g}{a\overline{\lambda}}\right)^{\mu/2}\exp\left(-\frac{\overline{\lambda}}{\pi}
\int_0^{+\infty}\frac{1}{q^2+\overline{\lambda}^2}\, 
\ln\left\lbrack 1+\frac{(1-s)}{q^\mu}\left(\frac{g}{a}\right)^\mu\right\rbrack\, dq\right)
 \ \ \ \ (g\rightarrow +\infty),
\end{eqnarray}
where we have used the asymptotic behavior\ (\ref{eq3.13}) valid for all $0<\mu\le 2$. Using\ (\ref{eqA4.5}) in\ (\ref{eqA4.3}) yields
\begin{eqnarray}\label{eqA4.6}
&&u(g\overline{x},s)\sim\frac{1}{2i\pi g}\left(\frac{g}{a}\right)^{\mu/2} \\
&&\times\int_{\cal{L}}
d\overline{\lambda}\, \frac{\rm{e}^{\overline{\lambda}\overline{x}}}
{\overline{\lambda}^{\mu/2}}\, 
\exp\left(-\frac{\overline{\lambda}}{\pi}
\int_0^{+\infty}\frac{1}{q^2+\overline{\lambda}^2}\, 
\ln\left\lbrack 1+\frac{(1-s)}{q^\mu}\left(\frac{g}{a}\right)^\mu\right\rbrack\, dq\right)
 \ \ \ \ (g\rightarrow +\infty), \nonumber
\end{eqnarray}
which, together with the algebraic tail\ (\ref{eq4.1.1.1}), gives
\begin{eqnarray}\label{eqA4.7}
&&I_1(g,s)\sim\frac{C_\alpha d^\alpha s}{g^{\alpha +1}}
\left(\frac{g}{a}\right)^{\mu/2}
\int_0^{+\infty}\frac{d\overline{x}}{(1+\overline{x})^{\alpha +1}} \\
&&\times\int_{\cal{L}}
\frac{d\overline{\lambda}}{2i\pi}\, \frac{\rm{e}^{\overline{\lambda}\overline{x}}}
{\overline{\lambda}^{\mu/2}}\, 
\exp\left(-\frac{\overline{\lambda}}{\pi}
\int_0^{+\infty}\frac{1}{q^2+\overline{\lambda}^2}\, 
\ln\left\lbrack 1+\frac{(1-s)}{q^\mu}\left(\frac{g}{a}\right)^\mu\right\rbrack\, dq\right)
 \ \ \ \ (g\rightarrow +\infty). \nonumber
\end{eqnarray}
Thus, in the large $g$ and $l$ limit one has
\begin{eqnarray}\label{eqA4.8}
&&\frac{1}{2i\pi}\oint\frac{I_1(g,s)}{s^{l+1}}\, ds\sim
\frac{C_\alpha d^\alpha}{lg^{\alpha +1}}
\left(\frac{g}{a}\right)^{\mu/2}\int_{\cal{L}}\frac{d\overline{p}}{2i\pi}{\rm e}^{\overline{p}}
\int_0^{+\infty}\frac{d\overline{x}}{(1+\overline{x})^{\alpha +1}} \\
&&\times\int_{\cal{L}}
\frac{d\overline{\lambda}}{2i\pi}\, \frac{\rm{e}^{\overline{\lambda}\overline{x}}}
{\overline{\lambda}^{\mu/2}}\, 
\exp\left(-\frac{\overline{\lambda}}{\pi}
\int_0^{+\infty}\frac{1}{q^2+\overline{\lambda}^2}\, 
\ln\left\lbrack 1+\frac{\overline{p}}{lq^\mu}\left(\frac{g}{a}\right)^\mu\right\rbrack\, dq\right)
 \ \ \ \ (g\ {\rm and}\ l\rightarrow +\infty), \nonumber
\end{eqnarray}
where we have made the change of variables $s=\exp(-\overline{p}/l)$ and use the fact that, in the $l\rightarrow +\infty$ limit, only the vicinity of $s=1$ contributes to the $s$-integral. Finally, using\ (\ref{eqA4.8}) and the large $g$ behavior\ (\ref{eq4.1.1.2}) of $I_2(g)$ in\ (\ref{eqA4.1}), one obtains the scaling form, [we recall that $\mu$ is related to $\alpha$ by $\mu =\min(2,\alpha)$],
\begin{equation}\label{eqA4.9}
p(g,l)\sim\frac{d^{2\alpha}}{g^{1+2\alpha}}F_{\alpha}\left(\frac{a^\mu l}{g^\mu}\right)
 \ \ \ \ (l,g\rightarrow +\infty),
 \end{equation}
 where
 \begin{eqnarray}\label{eqA4.10}
F_{\alpha}(y)&=&\frac{\sqrt{\pi}B_{\alpha}}{y}
\int_{\cal{L}}\frac{d\overline{p}}{2i\pi}\, {\rm e}^{\overline{p}}
\int_0^{+\infty}\frac{d\overline{x}}{(1+\overline{x})^{\alpha +1}} \\
&\times&\int_{\cal{L}}
\frac{d\overline{\lambda}}{2i\pi}\, \frac{\rm{e}^{\overline{\lambda}\overline{x}}}
{\overline{\lambda}^{\mu/2}}\, 
\exp\left\lbrack-\frac{\overline{\lambda}}{\pi}
\int_0^{+\infty}\frac{1}{q^2+\overline{\lambda}^2}\, 
\ln\left(1+\frac{\overline{p}}{yq^\mu}\right)\, dq\right\rbrack , \nonumber
\end{eqnarray}
with
\begin{equation}\label{eqA4.11}
B_{\alpha}=\frac{C_\alpha^2\Gamma(\alpha-\mu/2)}{\sqrt{\pi}\Gamma(\alpha +1)}.
\end{equation}

To determine the small argument behavior of $F_{\alpha}(y)$ we make the change of variable $q\rightarrow\overline{\lambda}q$ and let $y\rightarrow 0$. It comes
\begin{eqnarray}\label{eqA4.12}
&&\exp\left\lbrack-\frac{1}{\pi}
\int_0^{+\infty}\frac{1}{1+q^2}\, 
\ln\left(1+\frac{\overline{p}}{y\overline{\lambda}^\mu q^\mu}\right)\, dq\right\rbrack
\sim\exp\left\lbrack-\frac{1}{\pi}
\int_0^{+\infty}\frac{1}{1+q^2}\, 
\ln\left(\frac{\overline{p}}{y\overline{\lambda}^\mu q^\mu}\right)\, dq\right\rbrack
\nonumber \\
&&=\exp\left\lbrack\frac{1}{\pi}
\ln\left(\frac{y\overline{\lambda}^\mu}{\overline{p}}\right)\int_0^{+\infty}\frac{dq}{1+q^2}
+\frac{\mu}{\pi}\int_0^{+\infty}\frac{\ln q}{1+q^2}\, dq\right\rbrack \nonumber \\
&&=\overline{\lambda}^{\mu/2}\, \sqrt{\frac{y}{\overline{p}}}
 \ \ \ \ (y\rightarrow 0),
\end{eqnarray}
where the first and second integrals over $q$ in the second line of\ (\ref{eqA4.12}) are respectively equal to $\pi /2$ and $0$. Injecting\ (\ref{eqA4.12}) into the right-hand side of\ (\ref{eqA4.10}), one finds
\begin{eqnarray}\label{eqA4.13}
F_{\alpha}(y)&\sim&\frac{\sqrt{\pi}B_{\alpha}}{\sqrt{y}}
\left(\int_{\cal{L}} \frac{{\rm e}^{\overline{p}}}{\sqrt{\overline{p}}}\, \frac{d\overline{p}}{2i\pi}\right)
\int_0^{+\infty}\left(\int_{\cal{L}}\rm{e}^{\overline{\lambda}\overline{x}}
\, \frac{d\overline{\lambda}}{2i\pi}\right)\, 
\frac{d\overline{x}}{(1+\overline{x})^{\alpha +1}} \nonumber \\
&=&\frac{B_{\alpha}}{\sqrt{y}}
 \ \ \ \ (y\rightarrow 0),
\end{eqnarray}
where the integrals over $\overline{p}$ and $\overline{\lambda}$ are respectively equal to $1/\sqrt{\pi}$ and $\delta(\overline{x}-0^+)$.

The large argument behavior of $F_{\alpha}(y)$ is obtained along the same line as the analysis of the behavior of $\phi(\lambda ,s)$ near $s=1$ in Section\ \ref{sec4} and Appendix\ \ref{app3}. Skipping the details, one finds
\begin{eqnarray}\label{eqA4.14}
F_{\alpha}(y)&\sim& -\frac{\sqrt{\pi}\, B_{\alpha}}{\sin(\pi/\mu)}\frac{1}{y^{1+1/\mu}}
\left(\int_{\cal{L}}\overline{p}^{1/\mu}{\rm e}^{\overline{p}}\, \frac{d\overline{p}}{2i\pi}\right)
\int_0^{+\infty}\left(\int_{\cal{L}}\frac{\rm{e}^{\overline{\lambda}\overline{x}}}
{\overline{\lambda}^{1+\mu/2}}
\, \frac{d\overline{\lambda}}{2i\pi}\right)\, 
\frac{d\overline{x}}{(1+\overline{x})^{\alpha +1}} \nonumber \\
&=& -\frac{\sqrt{\pi}\, B_{\alpha}}{\sin(\pi/\mu)\Gamma(-1/\mu)\Gamma(\mu/2 +1)}
\frac{1}{y^{1+1/\mu}}\int_0^{+\infty}\frac{\overline{x}^{\mu/2}}{(1+\overline{x})^{\alpha +1}}
\, d\overline{x} \nonumber \\
&=& -\frac{\sqrt{\pi}\, B_{\alpha}\, \Gamma(\alpha -\mu/2)}
{\sin(\pi/\mu)\Gamma(-1/\mu)\Gamma(\alpha +1)}
\frac{1}{y^{1+1/\mu}} \ \ \ \ (y\rightarrow +\infty),
\end{eqnarray}
if $1/\mu$ is not an integer, and
\begin{eqnarray}\label{eqA4.15}
F_{\alpha}(y)&\sim& (-1)^{1+1/\mu}\frac{B_{\alpha}}{\sqrt{\pi}}
\left(\int_{\cal{L}}\overline{p}^{1/\mu}\ln(\overline{p})\, {\rm e}^{\overline{p} y}
\, \frac{d\overline{p}}{2i\pi}\right)
\int_0^{+\infty}\left(\int_{\cal{L}}\frac{\rm{e}^{\overline{\lambda}\overline{x}}}
{\overline{\lambda}^{1+\mu/2}}
\, \frac{d\overline{\lambda}}{2i\pi}\right)\, 
\frac{d\overline{x}}{(1+\overline{x})^{\alpha +1}} \nonumber \\
&=&\frac{B_{\alpha}\, \Gamma(1+1/\mu)}{\sqrt{\pi}\, \Gamma(\mu/2 +1)}
\frac{1}{y^{1+1/\mu}}\int_0^{+\infty}\frac{\overline{x}^{\mu/2}}{(1+\overline{x})^{\alpha +1}}
\, d\overline{x} \nonumber \\
&=& \frac{B_{\alpha}\, \Gamma(1+1/\mu)\Gamma(\alpha -\mu/2)}
{\sqrt{\pi}\, \Gamma(\alpha +1)}
\frac{1}{y^{1+1/\mu}} \ \ \ \ (y\rightarrow +\infty),
\end{eqnarray}
if $1/\mu$ is an integer, (where we have made the change of variable $\overline{p}\rightarrow y\overline{p}$). Using the reflection formula $\Gamma(-z)\Gamma(z+1)=-\pi/\sin(\pi z)$ in\ (\ref{eqA4.14}), substituting for $B_{\alpha}$ its expression\ (\ref{eqA4.11}), and writing
\begin{equation}
A_{\alpha}=\frac{C_\alpha^2 \Gamma(1+1/\mu)\Gamma(\alpha -\mu/2)^2}
{\pi\Gamma(\alpha +1)^2},
\end{equation}
one finally obtains
\begin{equation}
F_{\alpha}(y)\sim\frac{A_{\alpha}}{y^{1+1/\mu}}
 \ \ \ \ (y\rightarrow +\infty),
\end{equation}
for all $0<\mu\le 2$. [We recall that $\mu$ is related to $\alpha$ by $\mu =\min(2,\alpha)$ with $0<\alpha\ne 2$].
%
%

%
%

\begin{thebibliography}{1}
\bibitem{note1} To each walk with $l>0$ we can associate a walk with $l<0$ and the same probability by taking $x(n)$ as a new origin and reversing the direction of time, and conversely.
\bibitem{note2} Strictly speaking, we make this change of variable for real $\lambda >0$ and analytically continue the final result of the calculation to its analyticity domain in the complex plane. This remark applies throughout the article each time we make a change of variable involving some power of $\lambda$.
\bibitem{note3} This change of variable is to be understood in accordance with\ \cite{note2} in which $\lambda$ is replaced with $(1-s)$.
\end{thebibliography}
\end{document}